\documentclass[letterpaper,twocolumn,10pt]{article}
\usepackage{usenix2019_v3}

\usepackage{tikz}
\usepackage{amsmath}
\usepackage{mathptmx} 
\usepackage{fancyhdr}
\usepackage[normalem]{ulem}

\usepackage{graphicx}
\usepackage{graphics}
\usepackage{epsfig}
\usepackage{multicol}
\usepackage[shortlabels]{enumitem}

\usepackage{color}
\usepackage{lipsum}
\usepackage{verbatim}
\usepackage{alltt}
\usepackage{amsmath}
\usepackage{amssymb}
\usepackage{float}
\usepackage[]{algorithm2e}
\usepackage{balance}
\usepackage{tikz}
\usepackage{calc}
\usepackage{multirow}
\usepackage{pifont}
\usepackage{adjustbox}
\usepackage{array}
\usepackage{slashbox}
\usepackage{hhline}
\usepackage{pifont}
\usepackage[title]{appendix}

\newcommand{\ignore}[1]{}
\newcommand{\mirage}{Mirage\xspace}

\newcommand*\circled[1]{\tikz[baseline=(char.base)]{%
            \node[shape=circle,fill=black,draw,inner sep=0.5pt] (char) {\color{white}\fontfamily{phv}\selectfont\textbf{#1}};}}

\newcommand{\cmark}{\ding{51}}%
\newcommand{\xmark}{\ding{55}}%

\usepackage[firstpage]{draftwatermark}
\SetWatermarkText{\hspace*{6in}\raisebox{8.3in}{\includegraphics{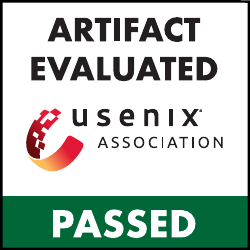}}}
\SetWatermarkAngle{0}

\begin{document}

\date{}
\title{\vspace{-0.3in}MIRAGE: Mitigating Conflict-Based Cache Attacks \\with a Practical Fully-Associative Design\vspace{-0.15in}}

\author{
{\rm Gururaj\ Saileshwar}\\
gururaj.s@gatech.edu \\
Georgia Institute of Technology
\and
{\rm Moinuddin Qureshi}\\
moin@gatech.edu \\
Georgia Institute of Technology
}

\vspace{0.3in}

\maketitle

\begin{abstract}

Shared caches in processors are vulnerable to conflict-based side-channel attacks, whereby an attacker can monitor the access pattern of a victim by evicting victim cache lines  using cache-set conflicts. Recent mitigations propose randomized mapping of addresses to cache lines, to obfuscate the locations of set-conflicts. However, these are vulnerable to newer attack algorithms that discover conflicting sets of addresses despite such mitigations, because these designs select candidates for eviction from a small set of conflicting lines.




This paper presents \textit{\mirage}, a practical design for a fully associative cache, wherein eviction candidates are selected randomly from among all the lines resident in the cache, to be immune to set-conflicts. A key challenge  in enabling such a design for large shared caches  (containing tens of thousands of resident cache lines) is managing the complexity of cache-lookup, as a naive design can require searching through all the resident lines. Mirage achieves full-associativity while retaining practical set-associative lookups by decoupling placement and replacement, using pointer-based indirection from tag-store to data-store to allow a newly installed address to globally evict the data of any random resident line. To eliminate set-conflicts, Mirage provisions extra invalid tags in a skewed-associative tag-store design where lines can be installed without set-conflict, along with a load-aware skew-selection policy that guarantees the availability of sets with invalid tags. Our analysis shows \mirage provides the global eviction property of a fully-associative cache throughout system lifetime (violations of full-associativity, i.e. set-conflicts, occur less than once in $10^{4}$ to $10^{17}$  years), thus offering a principled defense against any eviction-set discovery and any potential conflict based attacks. Mirage incurs limited slowdown (2\%) and 17--20\% extra storage compared to a non-secure cache.
\end{abstract}

\vspace{-0.1 in}
\section{Introduction}

Ensuring effective data security and privacy in the context of hardware side channels\ignore{, while retaining performance} is a challenge\ignore{ for designing computing hardware}. Performance-critical hardware components such as last-level caches (LLC) are often designed as shared resources to maximize utilization.  When a sensitive application shares the LLC with a malicious application running simultaneously on a different core, cache side channels can leak sensitive information. Such cache attacks have been shown to leak sensitive data like encryption keys~\cite{Bernstein} and  user data in the cloud~\cite{GetOffMyCloud}. Set-conflict based cache attacks (e.g. \textit{Prime+Probe}~\cite{PrimeProbe}) are particularly potent as they do not require any shared memory between the victim and the spy and exploit the set-associative design of conventional caches. Such designs map addresses to only a small group of cache locations called a \textit{set}, to enable efficient cache lookup. If the addresses of both the victim and the attacker map to the same set, then they can evict each other from the cache (such an episode is called a \textit{set-conflict}) -- the attacker uses such evictions to monitor the access pattern of the victim. 




\begin{figure*}[htb] 
  	\centering
  \vspace{-0.09 in}
	\includegraphics[width=6.5in]{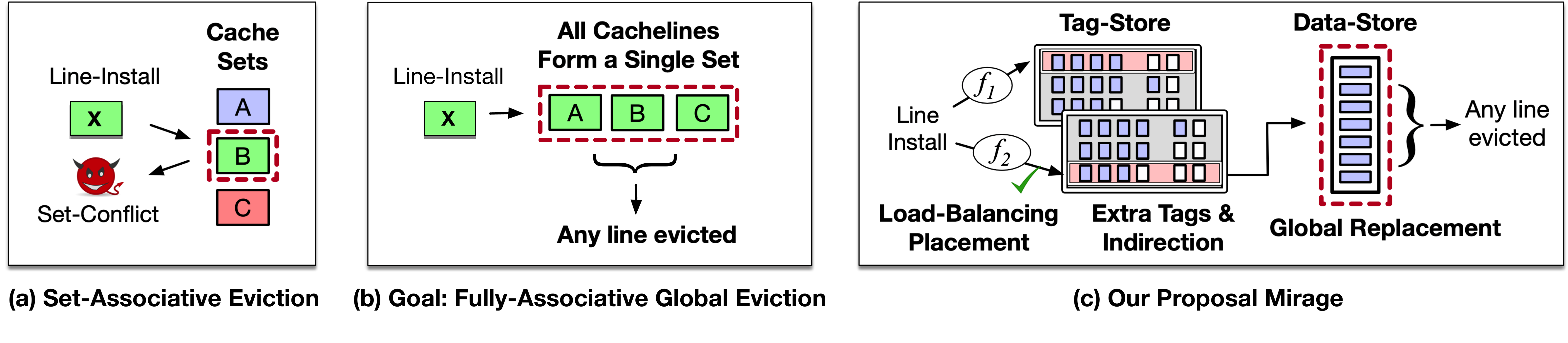}
     \vspace{-0.2 in}
  \caption{(a) Traditional LLCs have set-associative evictions (SAE), which leaks information to a spy. (b) Desired abstraction:  Global Evictions (GLE) on misses that avoid set conflicts. (c) Our proposal, Mirage, enables global evictions practically with: (1) Indirection from tag-store to the data-store,  (2) Skewed-Associative tag-store with extra tags, and (3)~Placement of lines with load-balancing that guarantees the availability of sets with invalid tags and eliminates SAE. }

  \vspace{-0.175in}
	\label{fig:intro} 
	
\end{figure*}

Recent proposals for \textit{Randomized LLCs}~\cite{micro18:CEASER,ScatterCache,isca19:SkewedCEASER,PhantomCache} attempt to mitigate set-conflict-based attacks by randomizing the locations of cachelines, i.e. addresses resident in the cache. By making the address-to-set mapping randomized and unpredictable to an adversary, these designs attempt to obfuscate the locations of the lines that are evicted. However, such defenses continue to select cachelines for eviction from a small number of locations in the cache (equal to the cache associativity), as shown in Figure~\ref{fig:intro}(a), and thus set-conflicts continue to occur although their locations are obfuscated. Subsequent attacks ~\cite{sp19:FastEvictionSets,isca19:SkewedCEASER,KULeuven:ProbabilisticEvictionSets} have proposed efficient algorithms to discover a minimal \textit{eviction-set} (lines mapping to the same set as a target address, that can evict the target via set-conflicts) even in the presence of such defenses, rendering them ineffective. In this paper, we target the root cause of vulnerability to eviction-set discovery in prior defenses --  the limitation of selecting victims for eviction from a small subset of the cache (a few tens of lines), which allows an adversary, that observes evictions, to learn finite information about installed addresses.

Our goal is to eliminate set-conflicts and attacks that exploit them, with a cache that has the property of  {\em global evictions}, i.e the victims for eviction are chosen (randomly) from among \textit{all} the lines in the cache. With global evictions, any line resident in the cache can get evicted when a new address is installed into the cache; all cachelines belong to a single set as shown in Figure~\ref{fig:intro}(b). Hence, an adversary observing an eviction of its address gains no information about the installed address. 

A fully associative cache design, where an address can map to any location in the cache, naturally provides global evictions. However, the main challenge in adopting such a design for the LLC is ensuring practical cache lookup. As a line can reside in any cache location, a cache lookup can require searching through the entire LLC (containing tens of thousands of lines) and be much slower than even a memory access. \ignore{ In contrast, a set-associative design has efficient lookup, as it searches through only the lines within a set (typically 8 -- 32 lines).} Ideally, we want the security of a fully-associative design, but the practical lookup of a set-associative design.

To this end, we propose \textit{Mirage ({\underline M}ulti-{\underline I}ndex {\underline Ra}ndomized Cache with {\underline G}lobal {\underline E}victions)}.\ignore{, a design that provides the security of a fully associative design, while retaining efficient cache lookup.} The key insight in Mirage is the decoupling of \textit{placement} of a new line in the tag-store (where the metadata is stored, that determines the complexity of lookup), from the \textit{replacement} decisions (which locations should be evicted to free up capacity in the data-store). This allows the placement of the tag of the line in a small number of possible locations in the tag-store for efficient lookup, while selecting data victims globally from the entire data-store.






To enable global evictions, Mirage uses pointer-based indirection to associate tags with data-blocks and vice-versa (inspired by V-way Cache~\cite{Vway}) as shown in Figure~\ref{fig:intro}(c), unlike traditional caches that have an implicit mapping between the tag and data of a cacheline. Moreover, Mirage provisions extra invalid tags in each set of the tag-store at a modest storage cost (while retaining the same data-store capacity) and guarantees the availability of such invalid tags in each set with a high probability. Thus, when a new line is installed, an invalid tag can be allocated from the tag-store without requiring an eviction of a line from the same set. An eviction of a line is only required to free a data-block, which is selected randomly from all the lines in the data-store, providing global eviction.


It is essential to prevent the adversary from mapping several lines at a time to a specific set, to fully deplete the available tags in that set\ignore{, by knowing the address-to-set mapping}. On an install to such a fully-occupied set, the cache is forced to perform a {\em Set Associative Eviction (SAE)}, where a valid tag from the same set needs to be evicted to accommodate the incoming line. By observing such an SAE, an adversary can infer the address of the installed line causing the eviction, and eventually launch a set-conflict based attack.

To eliminate set-conflicts and SAE, and ensure all evictions are global evictions, Mirage first splits the tag store into two equal parts (skews), and uses a cryptographic hash function to randomize the line-to-set mapping within each skew, like prior skewed-associative designs for secure caches~\cite{isca19:SkewedCEASER,ScatterCache}. This allows a line the flexibility of mapping to two possible sets (one in each skew), in a manner unpredictable to the adversary. As both skews could have invalid tag-store entries, an important consideration is the skew-selection policy on a line-install\ignore{, that determines how invalid tags are distributed across sets}. Using a random skew-selection policy, such as in prior works~\cite{isca19:SkewedCEASER,ScatterCache}, results in an unbalanced distribution of invalid tags across sets, causing the episodes of SAE to continue to occur every few microseconds (a few thousand line installs). To promote a balanced distribution of invalid tags across sets, Mirage employs a load-aware skew selection policy (inspired by load-aware hashing~\cite{Powerof2ChoicesSurvey,BalancedAllocations}), that chooses the skew with the most invalid tag-entries in the given set.  With this policy, Mirage guarantees an invalid tag is always available for an incoming line for system lifetime, thus eliminating SAE.


For an LLC with 2MB/core capacity and 16-ways in the baseline, Mirage provisions 75\% extra tags, and has two skews, each containing 14-ways of tag-store entries. Our analysis shows that such a design encounters SAE once per $10^{17}$ years, providing the global eviction property and an illusion of a fully associative cache virtually throughout system lifetime.

If Mirage is implemented with fewer than 75\% extra tags, the probability of an SAE increases as the likelihood that the tag entries in both skews are all valid increases.  
To avoid an SAE in such cases, we propose an optimization that relocates an evicted tag to its alternative set that is likely to have invalid tags with high probability (note that each address maps to two sets, one in each skew).  Mirage equipped with such {\em Cuckoo Relocation} (inspired from cuckoo hashing~\cite{cuckoo_hashing}),  ensures an SAE occurs once every 22,000 years, with 50\% extra tags\ignore{, while reducing the extra tags from 75\% to 50\%}.



\ignore{
\begin{figure}[htb] 
\vspace{-0.1in}
  	\centering
   	\includegraphics[width=3.3in]{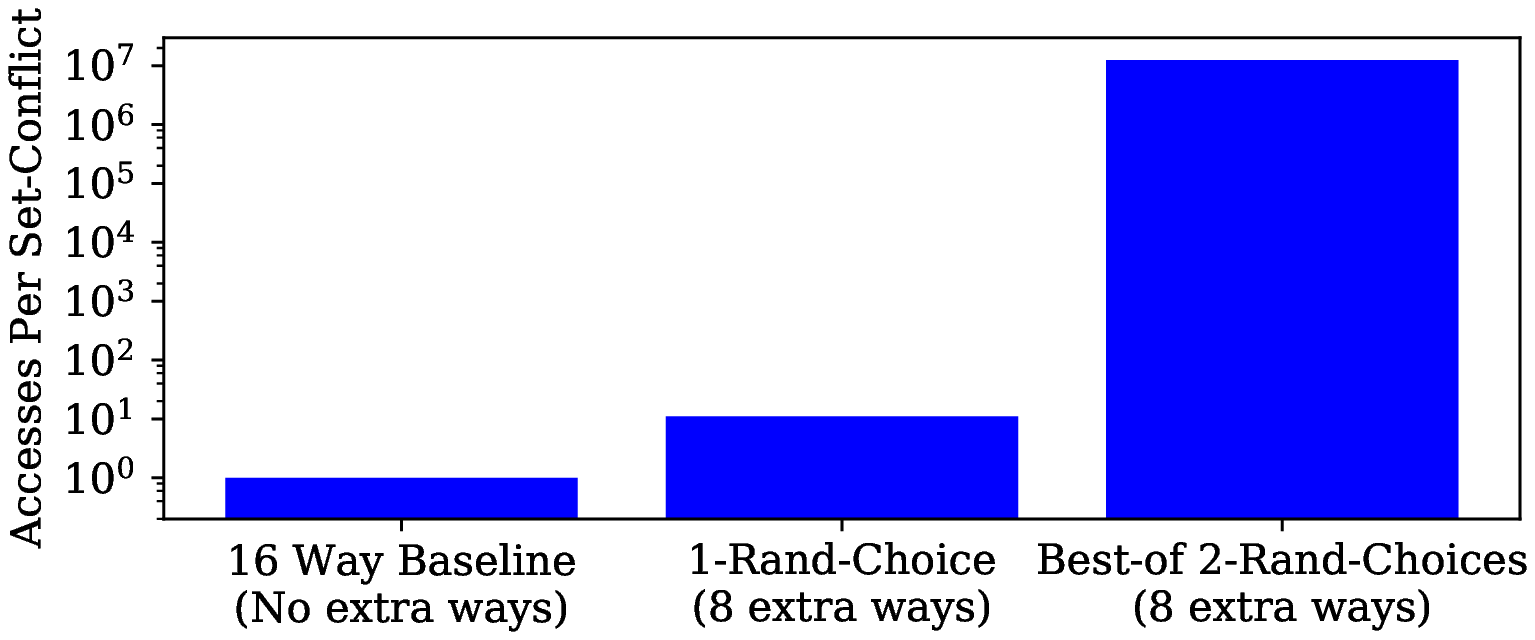}
    \vspace{-0.275 in}
	\caption{Accesses per set-conflict in a buckets and balls simulation of a 2MB cache over 100 million accesses.}
	\label{fig:introBnB}
\end{figure}
}

\vspace{0.05in}
\noindent Overall, this paper makes the following contributions:
\begin{enumerate}
  \setlength\itemsep{0.02in}
    \item We observe that conflict-based cache attacks can be mitigated by having global eviction that considers all the lines for eviction. For practical adoption, our goal is provide such a global eviction property without incurring significant latency for cache-lookup or power overhead. 
    

    \item We propose \textit{Mirage}, a practical way to get the global eviction benefits of a fully associative cache. Mirage uses indirection from tag-store to data-store, an intelligent tag store design, and a load balancing policy to ensure that the cache provides global evictions for system lifetime (set-associative evictions occur once in $10^{17}$ years).

    \item We propose {\em Mirage with Cuckoo Relocation}, whereby set-associative evictions in the tag store are mitigated by relocating a conflicting entry to an alternative location.\ignore{ in the other skew.  This ensures set-associative evictions occur once in 22,000 years while reducing the extra tags needed. } 
    
\end{enumerate}
\ignore{
\noindent \textbf{Benefits:} Mirage provides the illusion of a fully-associative cache with all evictions being randomly chosen globally from the entire cache, to eliminate conflict-based attacks. For a 16MB LLC, Mirage-v1 provisioned with 75\% extra tags over a 16-way design retains this illusion for $10^{17}$ years. On the other hand, Mirage-v2 provisioned with 50\% extra tags retains this illusion for 22,000 years. We analytically prove these bounds in  Section~\ref{sec:analytical_model}. These benefits are provided by Mirage while retaining practical cache lookup, with accesses to 24~--~28 ways over 2 sets required per cache lookup.
}



As Mirage requires extra tags and indirection, 
it incurs a modest storage overhead of 17\% to 20\% for a cache design with 64-byte linesize compared to a non-secure design.
Our evaluations show that
Mirage incurs a modest slowdown of 2\%, compared to  a non-secure set-associative baseline cache.
\section{Background and Motivation}
\subsection{Cache Design in Modern Processors}
Processor caches are typically organized at the granularity of 64-byte cache lines. A cache is typically divided into two structures -- the \textit{tag-store} and the \textit{data-store}. For each cacheline, the metadata used for identification (e.g. address, valid-bit, dirty-bit) is called the \textit{tag} and stored in the "tag-store", and there is a one-to-one mapping of the tag  with the \textit{data} of the line, which is stored in the "data-store". To enable efficient cache lookups, the tag-store is typically organized in a set-associative manner, where each address maps to a \textit{set} that is a group of contiguous locations within the tag-store, and each location within a set is called a \textit{way}. Each set consists of $w$ ways, typically in the range of 8 - 32 for  caches in modern processors ($w$ is also referred to as the cache associativity). As last-level caches (LLCs) are shared among multiple processor cores for performance, cachelines of different processes can contend for the limited space within a set, and evict each other from the cache -- such episodes of "set-conflicts" are exploited in side-channel attacks to evict victim cachelines.
\subsection{Threat Model}
We assume a threat model where the attacker and victim execute simultaneously on different physical cores sharing an LLC, that is inclusive of the L1/L2 caches private to each core. We focus on conflict-based cache side-channel attacks where the attacker causes set-conflicts to evict a victim's line and monitor the access pattern of the victim. Such attacks are potent as they do not require victim and attacker to access any shared memory. For simplicity, we assume no shared memory between victim and attacker, as existing solutions~\cite{ScatterCache} are effective at mitigating possible attacks on shared lines.\footnote{If the attacker and the victim have shared-memory, attacks such as Flush+Reload or Evict+Reload are possible. These can be mitigated by storing duplicate copies of shared-addresses, as proposed in Scatter-Cache~\cite{ScatterCache}\ignore{,and hence not our primary focus}. We discuss how our design incorporates this mitigation in Section~\ref{sec:FlushReload}.}


\subsection{Problem: Conflict-Based Cache Attacks}

\begin{figure}[htb] 
  	\centering
	\includegraphics[width=2.75 in]{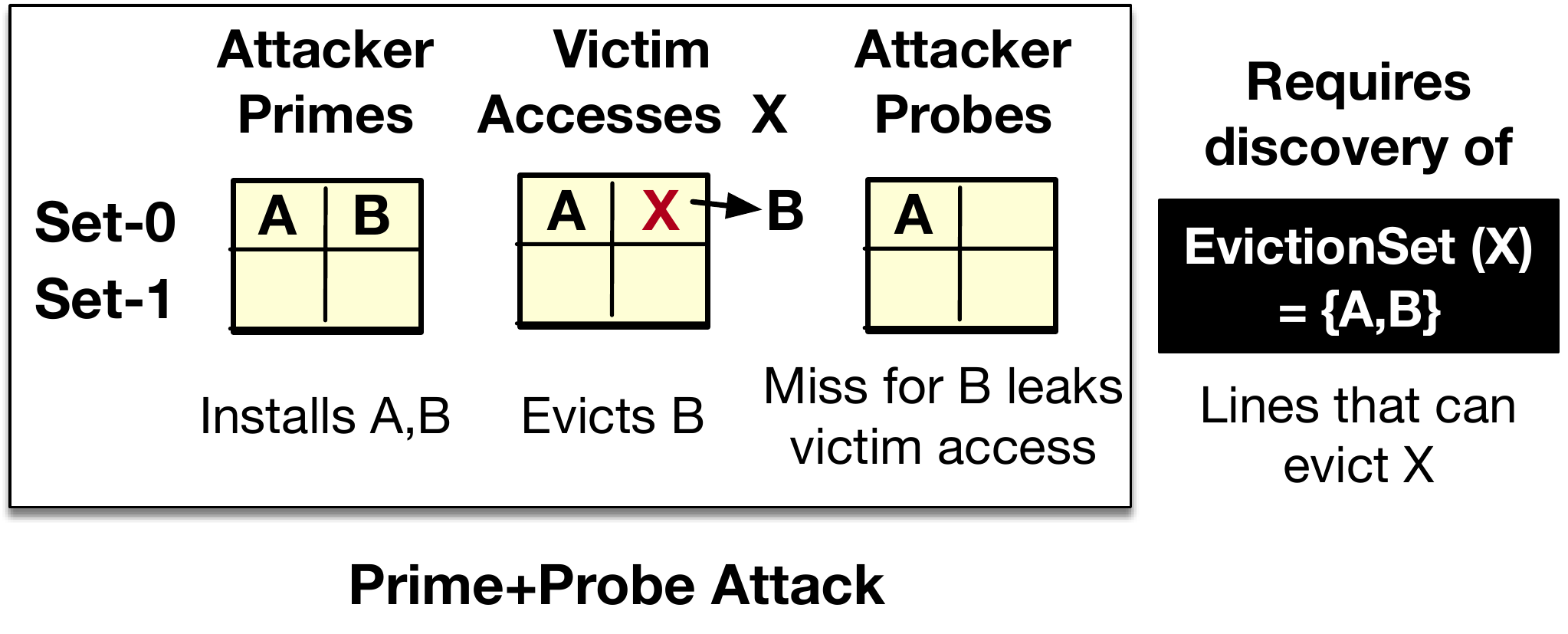}
    \vspace{-0.15in} 
	\caption{Example of Conflict-Based Attack (Prime+Probe).}
    \vspace{-0.1in} 
	\label{fig:primeprobe} 
\end{figure}

Without loss of generality, we describe the Prime+Probe attack~\cite{PrimeProbe} as an example of a conflict-based cache attack.  As shown in Figure~\ref{fig:primeprobe}, the attacker first primes a set with its addresses, then allows the victim to execute and evict an attacker line due to cache-conflicts. Later, the attacker probes the addresses to check if there is a miss, to infer that the victim accessed that set. \ignore{Alternately, in Evict+Time, the attacker learns if a victim accesses an address by first evicting the victim line from the cache using cache-conflicts, and then checking if the execution time of the victim increases due to the evicted address.} Prior attacks have monitored addresses accessed in AES T-table and RSA Square-Multiply Algorithms to leak secret keys~\cite{sp15:LLCAttacksPractical}, addresses accessed in DNN computations to leak DNN model parameters~\cite{CacheTelepathy}, etc. To launch such attacks, the attacker first needs to generate an \textit{eviction-set} for a victim address, i.e. a minimal set of addresses mapping to the same cache set as the victim address. \ignore{ As commercial processors use undocumented functions to map addresses to cache-sets, these attacks algorithmically discover an eviction-set for a victim-address}

\subsection{Recent Advances in Attacks and Defenses}
\label{sec:eviction_set}

Given how critical eviction-set discovery is for such attacks, recent defense works have proposed randomized caches to obfuscate the address to set mapping and make it harder to learn eviction sets. At the same time, recent attacks have continued to enable faster algorithms for eviction set discovery. We describe the key related works in this spirit and discuss the pitfalls of continuing with such an approach.


\vspace{0.12 in}

\noindent {\bf Move-1: Attack by Eviction Set Discovery in $O(n^2)$}

Typically, set-selection functions in caches are undocumented. A key work by Liu et al.~\cite{sp15:LLCAttacksPractical} proposed an algorithm to discover eviction-sets without the knowledge of the address to set mappings -- it tests and eliminates addresses one at a time, requiring $O(n^2)$ accesses to discover an eviction-set. 


\vspace{0.12 in}

\noindent{\bf Move-2: Defense via Encryption and Remapping}

\begin{figure}[htb] 
  	\centering
	\includegraphics[width=3.3in]{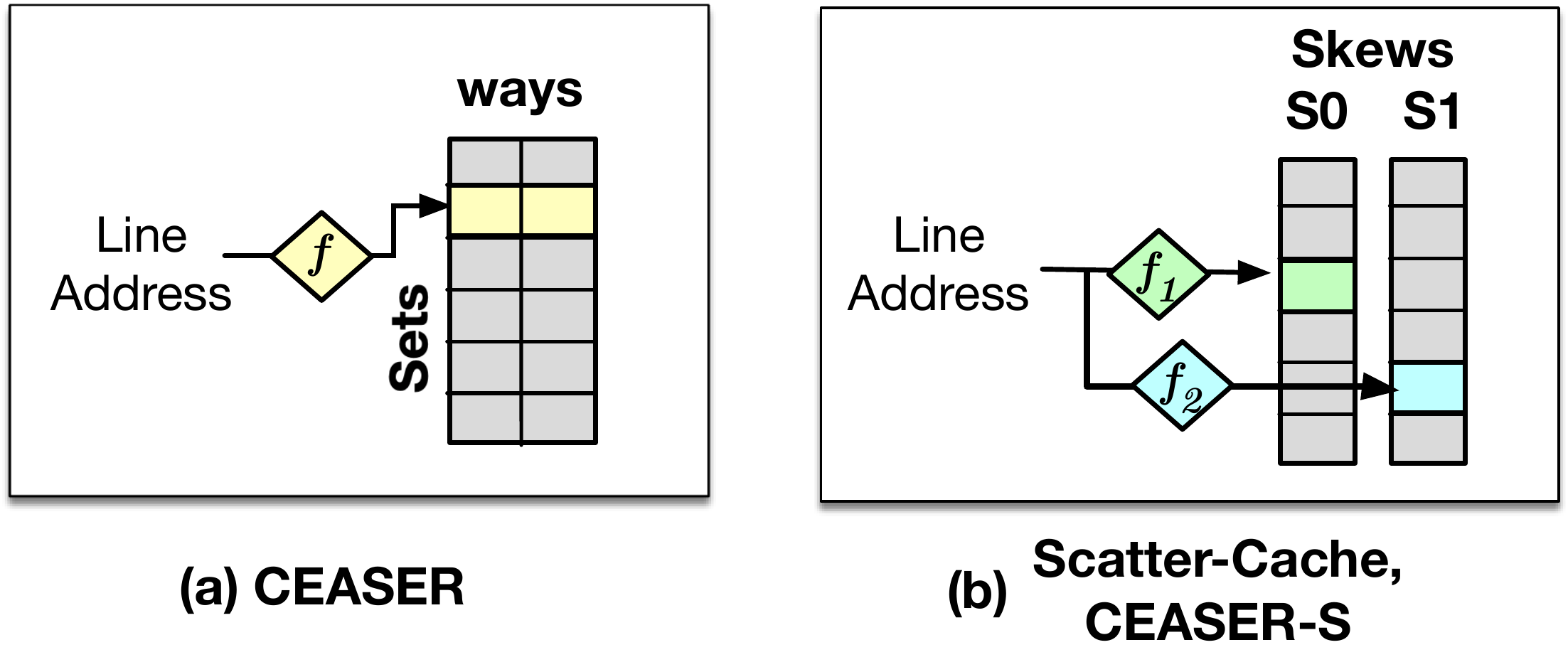}
    \vspace{-0.1in} 
	\caption{Recent Works on Randomized Caches}
    \vspace{-0.1in} 
	\label{fig:randcache} 
\end{figure}

CEASER~\cite{micro18:CEASER} (shown in Figure~\ref{fig:randcache}(a)) proposed randomizing the address to set mapping by accessing the cache with an encrypted line address. By enabling dynamic re-keying, it ensures that the mapping changes before an eviction-set can be discovered with an algorithm that requires $O(n^2)$ accesses.  

\begin{figure*}[htb] 
  	\centering
  \vspace{-0.09 in}
		\includegraphics[width=6in]{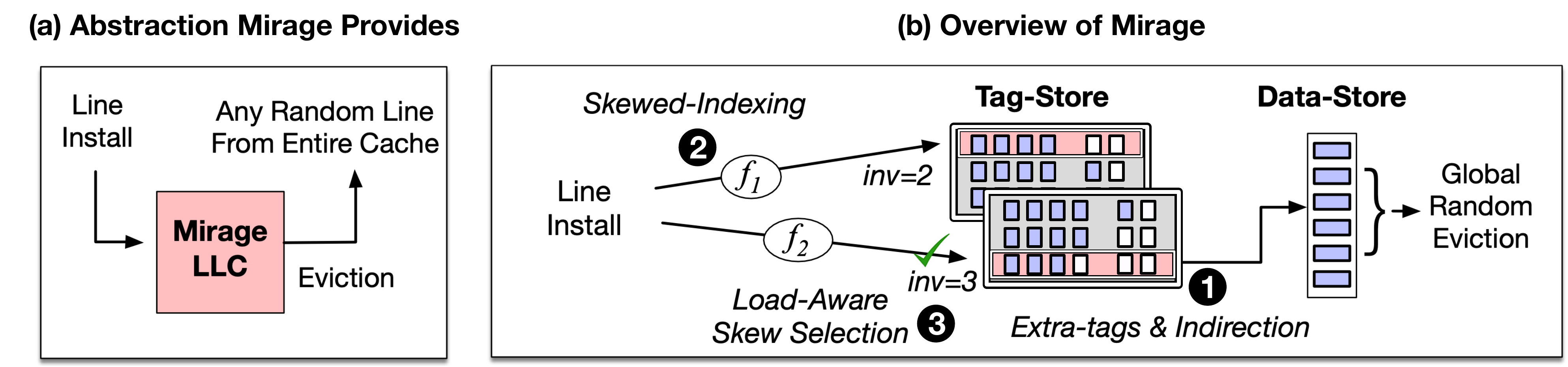}
      \vspace{-0.095 in}
    \caption{(a) Mirage provides the abstraction of a fully-associative design with globally random evictions. (b) It achieves this by using extra tags and indirection between tags and data-blocks, skewed-indexing, and load-aware skew-selection. \ignore{, to enable fully-associative evictions while retaining set-indexed cache lookup.}}

  \vspace{-0.125in}
	\label{fig:mirage_overview} 
	
\end{figure*}

\vspace{0.12 in}
\noindent{\bf Move-3: Attack by Eviction Set Discovery in $O(n)$}

Subsequent works~\cite{isca19:SkewedCEASER,sp19:FastEvictionSets} developed a faster algorithm that could discover eviction-sets in $O(n)$ accesses, by eliminating groups of lines from the set of potential candidates, rather than one line at a time. CEASER is unable to prevent eviction-set discovery with such faster algorithms.




\vspace{0.12 in}

\noindent {\bf Move-4: Defense via Skewed Associativity}

Scatter-Cache~\cite{ScatterCache} and CEASER-S~\cite{isca19:SkewedCEASER} adopt skewed associativity in addition to randomized mapping of addresses to sets, to further obfuscate the LLC evictions. As shown in Figure~\ref{fig:randcache}(b), such designs partition the cache across ways into multiple skews, with each skew having a different set-mapping and a new address is installed in a randomly selected skew. Such a design provides greater obfuscation as eviction sets get decided by the line to skew mapping as well. These designs were shown to be immune to faster eviction set discovery algorithms~\cite{isca19:SkewedCEASER,sp19:FastEvictionSets} that require $O(n)$ steps.


\noindent {\bf Move-5: Attack by Probabilistic Eviction Set Discovery}

A recent work~\cite{KULeuven:ProbabilisticEvictionSets} showed that faster eviction-set discovery in Scatter-Cache is possible with an intelligent choice of initial conditions, that boosts the probability of observing conflicts. This allows discovery of partial eviction-sets (lines that evict a target in a subset of the ways) within 140K accesses in Scatter-Cache, which can enable a conflict-based attack. 




\vspace{0.09 in}

\noindent {\bf Pitfalls:} \ignore{In essence, t}There is an interplay between the robustness of defenses and algorithms for eviction set discovery. The security of past defenses has hinged on obfuscation of eviction-sets. However, newer algorithms enabling faster eviction-set discovery continue to break such defenses. Ideally, we seek a defense that eliminates {\em Set-Associative Evictions (SAE)}, which are the root cause of the vulnerability, as they allow the adversary to learn eviction-sets. Eliminating SAE would not only safeguard against current algorithms for eviction set discovery but also against a hypothetical oracular algorithm that can learn an eviction-set after observing just a single conflict.

\subsection{Goal: A Practical Fully-Associative LLC}
As a principled defense against conflict-based attacks, we seek to design a cache that provides {\em Global Eviction (GLE)}, i.e. the eviction candidates are selected from among all of the addresses resident in the cache when new addresses are installed. Such a defense would eliminate SAE and be immune to eviction-set discovery, as evicted addresses are independent of the addresses installed and leak no information about installed addresses. \ignore{Eviction-set discovery is futile with such a design, as all the addresses in the cache belong to a single set.} While a fully-associative design provides global evictions, it incurs prohibitive latency and power overheads when adopted for an LLC.\footnote{Recent works propose fully-associative designs for a subset of the cache (Hybcache~\cite{Hybcache}) or for L1-Caches (RPCache~\cite{RPCache}, NewCache\cite{NewCache}). These approaches are impractical for LLCs (see  Section~\ref{sec:SecureCacheFullAssoc}).}
 The goal of our paper is to develop an LLC design that guarantees global evictions while retaining the practical lookup of a set-associative cache.

\newpage
\section{Full Associativity via MIRAGE}
\vspace{-0.05in}
To guarantee global evictions practically, we propose \textit{Mirage ({\underline M}ulti-{\underline I}ndex {\underline Ra}ndomized Cache with {\underline G}lobal {\underline E}victions)}. Mirage provides the abstraction of a fully associative cache with random replacement, as shown in Figure~\ref{fig:mirage_overview}(a), with the property that on a cache miss, a random line is evicted from among all resident lines in the cache. \ignore{, and allows the incoming line to place the content in that location.}This ensures the evicted victim is  independent of the incoming line and no subset of lines in the cache form an eviction set.

\ignore{
We propose Mirage, a cache design that provides an abstraction of a fully-associative cache with random replacement. This provides a guarantee that on any cache install, evictions are globally selected from the entire cache in a manner unrelated to addresses installed, as shown in Figure~\ref{fig:mirage_overview}(a). Thus, an eviction-set constructed with any algorithm, using a sequence of such global evictions, spans the entire cache making conflict-based attacks infeasible.
}

\vspace{-0.1in}
\subsection{ Overview of Mirage}
\vspace{-0.1in}

\ignore{To provide global evictions while preserving set-associative lookup,}Mirage has three key components, as shown in Figure~\ref{fig:mirage_overview}(b). First, it uses a cache organization that decouples tag and data location and uses indirection to link tag and data entries\ignore{similar to the V-way cache~\cite{Vway})} (\circled{1} in Figure~\ref{fig:mirage_overview}(b)). Provisioning extra invalid tags allows accommodating new lines in indexed sets without tag-conflicts, and indirection between tags and data-blocks allows victim-selection from the data-store in a global manner. Second, Mirage uses a tag-store design that splits the tag entries into two structures (skews) and accesses each of them with a different hashing function (\circled{2} in Figure~\ref{fig:mirage_overview}(b)). Finally, to maximize the likelihood of getting an invalid tag on cache-install, Mirage uses a load-balancing policy for skew-selection leveraging the "power of 2 choices"~\cite{Powerof2ChoicesSurvey} (\circled{3} in Figure~\ref{fig:mirage_overview}(b)), which ensures no SAE occurs in the system lifetime and all evictions are global. We describe each component next.

\vspace{-0.1in}
\subsection{Tag-to-Data Indirection and Extra Tags}
\label{sec:vway}

\begin{figure}[htb] 
  	\centering
	\includegraphics[width=3.2in]{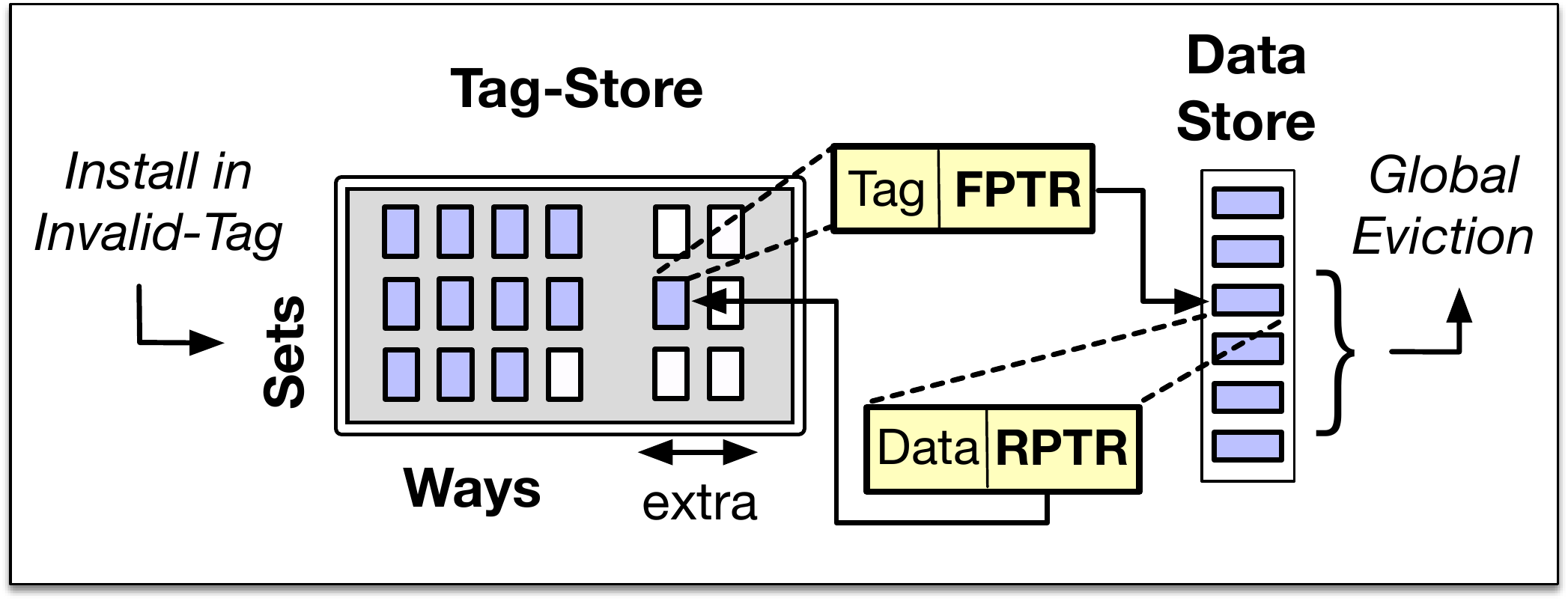}
    \vspace{-0.1in} 
	\caption{Overview of the cache substrate used by Mirage with indirection and extra tags (inspired by V-Way Cache).}
	\label{fig:vway_tags} 
\end{figure}

\textbf{V-way Cache Substrate}: Figure~\ref{fig:vway_tags} shows the tag and data store organization using pointer-based indirection in Mirage, which is inspired by the V-way cache~\cite{Vway}. V-way originally used this substrate to reduce LLC conflict-misses and improve performance. Here, the tag-store is over-provisioned to include extra invalid tags, which can accommodate the metadata of a new line without a set-associative eviction (SAE). Each tag-store entry has a forward pointer (FPTR) to allow it to map to an arbitrary data-store entry.\footnote{While indirection requires a cache lookup to serially access the tag and data entries, commercial processors~\cite{DEC,Itanium,AlbonesiSerialTagData} since the last two decades already employ such serial tag and data access for the LLC to save power (this allows the design to only access the data-way corresponding to the hit).} On a cache-miss, two types of evictions are possible: if the incoming line finds an invalid tag, a Global Eviction (GLE) is performed; else, an SAE is performed to invalidate a tag (and its corresponding data-entry) from the set where the new line is to be installed. 
On a GLE, V-way cache evicts a data entry intelligently selected from the entire data-store and also the corresponding tag identified using a reverse pointer (RPTR) stored with each data entry. In both cases, the RPTR of the invalidated data-entry is reset to invalid. This data-entry and the invalid tag in the original set are used by the incoming line. 


\textbf{Repurposing V-way Cache for Security}: Mirage adopts the V-way cache substrate with extra tags and indirection to enable GLE, but with an important modification: it ensures the data-entry victim on a GLE is selected randomly from the entire data-store (using a hardware PRNG) to ensure that it leaks no information. Despite this addition, the V-way cache substrate by itself is not secure, as it only reduces but does not eliminate SAE. For example, if an adversary has arbitrary control over the placement of new lines in specific sets, they can map a large number of lines to a certain set and deplete the extra invalid tags provisioned in that set. When a new (victim) line is to be installed to this set, the cache is then forced to evict a valid tag from the same set and incur an SAE.  Thus, an adversary who can discover the address to set mapping can force an SAE on each miss, making a design that naively adopts the V-way Cache approach vulnerable to the same attacks present in conventional set-associative caches.

\vspace{-0.25in}
\subsection{Skewed-Associative Tag-Store Design}
\label{sec:skewed_indexing}
To ensure GLE on each line install, Mirage reshapes the tag organization. \ignore{In a conventional set-associative tag-store, a new line is mapped to a single set and is forced to select a tag from only that set. If the invalid entries of that set are exhausted, it results in an SAE.} To allow an address to map to multiple sets in the tag store and increase the probability of obtaining an invalid tag, Mirage architects the tag-store as a skewed-associative structure~\cite{seznec_skew}. The tag store is split into two partitions or skews, and a different randomizing hash function is used to map addresses to sets in each skew. 
The hash function\footnote{The hash-function construction is similar to Scatter-Cache (SCv1)~\cite{ScatterCache}, where set-index bits are sliced from  a cipher-text encrypted using a plaintext of physical line-address concatenated with a Security-Domain-ID and the set-index  for each skew is computed using a different secret key.} to map addresses to sets is constructed using a 12-round PRINCE cipher~\cite{PRINCE}, which is a low-latency 64-bit block-cipher using 128-bit keys. Note that prior work~\cite{BRUTUS} used a reduced round version of PRINCE cipher for randomized cache indexing.

Unlike prior defenses using skewed-associativity~\cite{isca19:SkewedCEASER,ScatterCache}, each \ignore{tag-store} skew in Mirage contains invalid tags. Offering the flexibility for a new line to map to two sets (one in each skew) in the presence of invalid tags significantly increases the chance of finding an invalid tag in which it can be installed and avoiding an SAE.\ignore{, compared to a single choice offered in a conventional set-associative structure.} Moreover, the cryptographically generated address-to-set mapping ensures that the adversary (without knowing the secret key) cannot arbitrarily deplete these invalid tags within a set.

\vspace{-0.05in}
\subsection{Load-Aware Skew Selection}
\label{sec:random_choices}
Natural imbalance in usage of tags across sets can deplete invalid tags across sets and cause an SAE. On a line-install, the skew-selection policy, that decides the skew in which the line is installed, determines the distribution of invalid tags across sets. Prior works, including Scatter-Cache~\cite{ScatterCache} and CEASER-S~\cite{isca19:SkewedCEASER}, use random skew-selection, which randomly picks one of the two skews on a line-install. With invalid tags, this policy can result in imbalanced sets -- some with many invalid tags and others with none (that incur SAE). Our analysis, using a buckets-and-balls model we describe in Section~\ref{sec:bucketNballs}, indicates such a random skew-selection policy results in an SAE every few  misses (every 2600 misses with 6 extra ways/skew), and provides robustness only for microseconds.

To guarantee the availability of invalid tags across sets and eliminate SAE\ignore{ for system-lifetime}, Mirage uses a load-aware skew selection policy inspired by \textit{"Power of 2 Choices"}~\cite{Powerof2ChoicesSurvey,BalancedAllocations}, a load-balancing technique used in hash-tables. As indicated by \circled{3} in Figure~\ref{fig:mirage_overview}, this policy makes an intelligent choice between the two skews, installing the line in the skew where the indexed set has a higher number of invalid tags. In the case of a tie between the two sets, one of the two skews is randomly selected. With this policy, an SAE occurs only if the indexed sets in both skews do not have invalid tags, that is a rare occurrence as this policy actively promotes balanced usage of tags across sets.  Table~\ref{table:overflows_choice_indexing} shows the rate of SAE for Mirage  with load-aware skew selection policy, as the number of extra tags per skew is increased from 0 to 6. Mirage with 14-ways per skew (75\% extra tags) encounters an SAE once in  $10^{34}$ cache-installs, or equivalently  $10^{17}$ years, ensuring no SAE throughout the system lifetime.\ignore{\footnote{Prior work~\cite{BalancedAllocations} similarly shows that load-aware selection in hash-tables makes the max load/entry exponentially less compared to random selection.}} We derive these bounds analytically in Section~\ref{sec:analytical_model}.  



\begin{table}[htb]
  \begin{center}
\vspace{-0.2in}

   \renewcommand{\arraystretch}{1}
      \setlength{\tabcolsep}{4 pt}
    \caption{Frequency of Set-Associative Eviction (SAE) in Mirage as number of extra ways-per-skew is increased  (assuming 16-MB LLC with 16-ways in the baseline and 1ns per install)}
    \vspace{0.075in}
    \begin{small}
        \resizebox{2.8in}{!}{
    \ignore{
      \begin{tabular}{|c|c|c|c|c|c|c|} \hline
\textbf{Number of Extra Ways/Skew}& & 1 & 2 & 3 & 4 & 5& 6   \\ \hline \hline
      \textbf{LLC-Installs per SAE} & 4 & 60 & 800 &160mn \\ \hline
      \textbf{Time per SAE}  & $10^{34}$ & $10^{17}$ years  \\ \hline 
      \end{tabular}
}      

      \begin{tabular}{|c|c|c|} \hline
\textbf{Ways in each Skew}& \multirow{2}{*}{\bf Installs per SAE} & \multirow{2}{*}{\bf Time per SAE}   \\ 
\textbf{ (Base + Extra)} && \\ \hline \hline
        8 + 0 & 1 & 1 ns \\ \hline
        8 + 1 & 4 & 4 ns \\ \hline
        8 + 2 & 60 & 60 ns \\ \hline
        8 + 3 & 8000 & 8 us \\ \hline
        8 + 4 & $2\times 10^8$  & 0.16 s \\ \hline
        8 + 5 & $7 \times 10^{16}$ & 2 years \\ \hline
        8 + 6 (default Mirage) & $10^{34}$ & $10^{17}$ years \\ \hline
        
      \end{tabular}
      }
    \end{small}   
\label{table:overflows_choice_indexing}
\end{center}
\vspace{-0.55in}
\end{table}

\newpage
\section{Security Analysis of Mirage}
In this section, we analyze set-conflict-based attacks in a setting where the attacker and the victim do not have shared memory (shared-memory attacks are analyzed in Section~\ref{sec:FlushReload}). All existing set-conflict based attacks, such as Prime+Probe~\cite{PrimeProbe}, Prime+Abort~\cite{PrimeAbort}, Evict+Time~\cite{PrimeProbe}, etc. exploit eviction-sets to surgically evict targeted victim-addresses, and all eviction-set discovery algorithms require the attacker to observe evictions dependent on the addresses accessed by the victim. In Mirage, two types of evictions are possible -- a global eviction, where the eviction candidate is selected randomly from all the lines in the data-store, that leak no information about installed addresses\ignore{(assuming the hardware PRNG, which decides the candidates for global eviction, has sufficient entropy)}; or a set-associative eviction (SAE), where the eviction candidate is selected from the same set as the installed line due to a tag-conflict, that leaks information. To justify how Mirage eliminates conflict-based attacks, in this section we estimate the rate of SAE and reason that even a single SAE is unlikely to occur in system-lifetime. 

\vspace{0.07in}
Our security analysis makes the following assumptions:
\vspace{-0.04in}
\begin{enumerate}
    \setlength{\itemsep}{2pt}%
    \setlength{\parskip}{2pt}%

  
  \item \textbf{Set-index derivation functions are perfectly random and the keys are secret.} This ensures the addresses are uniformly mapped to cache-sets, in a manner unknown to the adversary, so that they cannot directly induce SAE. Also, the mappings in different skews (generated with different keys) are assumed to be independent, as required for the power of 2-choices load-balancing.
  
  
  \item \textbf{Even a single SAE is sufficient to break the security.} The number of accesses required to construct an eviction-set has reduced due to recent advances, with the state-of-the-art~\cite{sp15:LLCAttacksPractical,isca19:SkewedCEASER,sp19:FastEvictionSets} requiring at least a few hundred SAE to construct eviction-sets. To potentially mitigate even future advances in eviction-set discovery, we consider a powerful hypothetical adversary that can construct an eviction-set with just a single SAE (the theoretical minimum), unlike previous defenses~\cite{micro18:CEASER,isca19:SkewedCEASER,ScatterCache} that only consider existing eviction-set discovery algorithms. 
  
\end{enumerate}
%

\subsection{Bucket-And-Balls Model}
\label{sec:bucketNballs}
\begin{figure}[htb] 
  	\centering
	\includegraphics[width=3.5in]{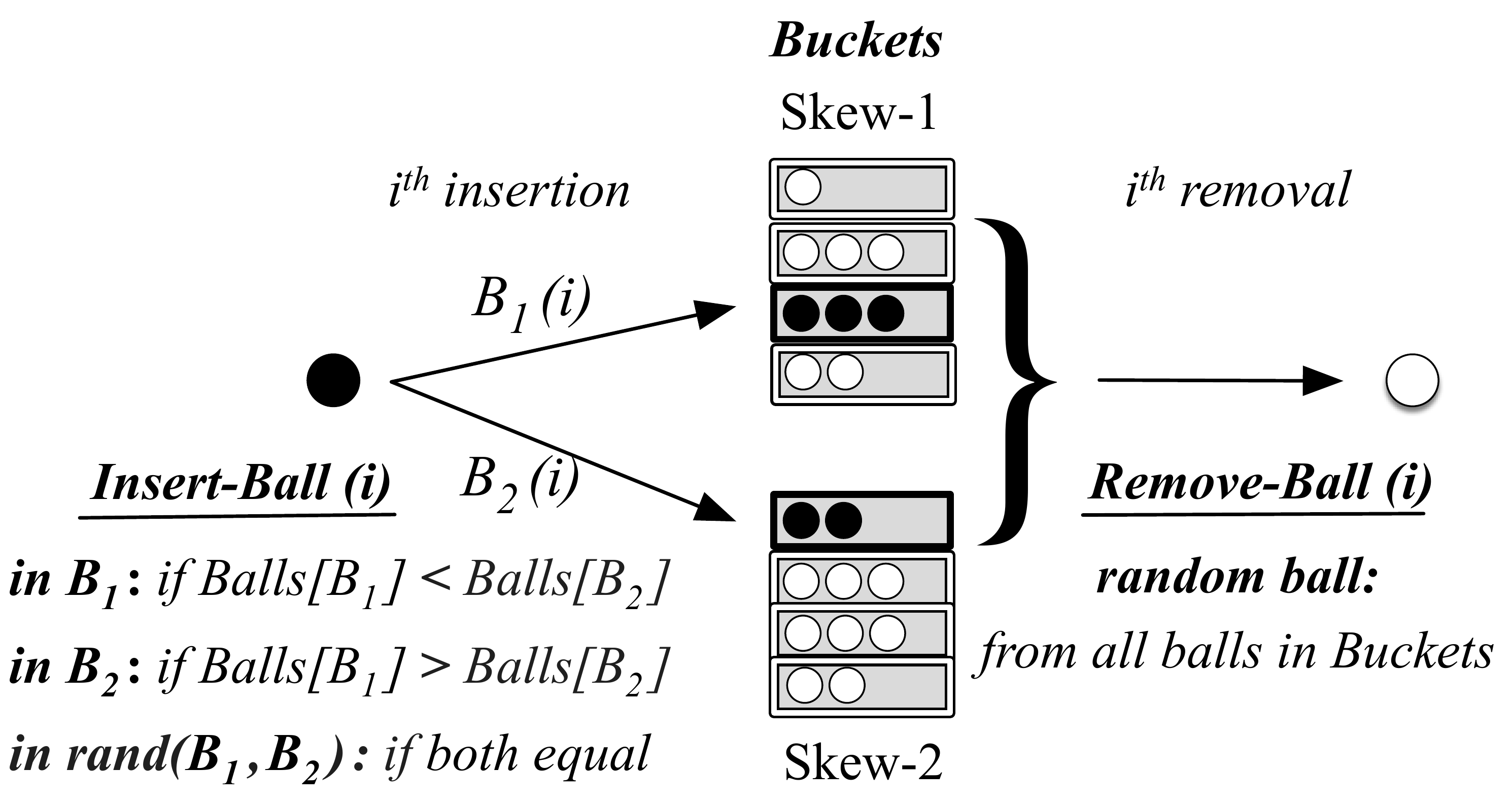}
    \vspace{-0.175in} 
	\caption{Buckets-and-balls model for Mirage with 32K buckets (divided into 2 skews), holding 256K balls in total to model a 16MB cache. The bucket capacity is varied from 8-to-14 to model 8-to-14 ways per skew in Mirage. }
	\label{fig:bucketNballs} 
\end{figure}

To estimate the rate of SAE, we model the operation of Mirage as a buckets-and-balls problem, as shown in Figure~\ref{fig:bucketNballs}. Here each bucket models a cache-set and each ball throw represents a new address installed into the cache. Each ball picks from 2 randomly chosen buckets, one from each skew, and is installed in the bucket with more free capacity, modeling the skew-selection in Mirage. If both buckets have the same number of balls, one of the two buckets is randomly picked.\footnote{A biased tie-breaking policy~\cite{LoadBalancing:AlwaysGoLeft} that always picks Skew-1 on ties further reduces the frequency of bucket-spills by few orders of magnitude compared to random tie-breaks. However, to keep our analysis simple, we use a random tie-breaking policy.} If both buckets are full, an insertion will cause a \textit{bucket spill}, equivalent to an SAE in Mirage. Otherwise, on every ball throw, we randomly remove a ball from among all the balls in buckets to model Global Eviction. The parameters of our model are shown in Table~\ref{table:BBconfig}.  We initialize the buckets by inserting as many balls as cache capacity (in number of lines) and then perform 10 trillion ball insertions and removals to measure the frequency of bucket spills (equivalent to SAE). Note that having fewer lines in the cache than the capacity is detrimental to an attacker, as the probability of a spill would be lower; so we model the best-case scenario for the attacker. 

\vspace{-0.1in}
\begin{table}[h]
  \begin{center}
    \begin{small}
      \caption{Parameters for Buckets and Balls Modeling }
       \vspace{0.1in}
        \renewcommand{\arraystretch}{1.15}

      \begin{tabular}{|l|l|} \hline

\bf Buckets and Balls Model & \bf Mirage Design \\ \hline
Balls - 256K & Cache Size - 16 MB \\
Buckets/Skew - 16K & Sets/Skew - 16K \\
Skews - 2 & Skews - 2 \\ 
Avg Balls/Bucket - 8 & Avg Data-Lines Per Set - 8 \\
Bucket Capacity - 8 to 14 & Ways Per Skew - 8 to 14  \\ \hline
      \end{tabular}
      \label{table:BBconfig}
 \end{small}
\end{center}
  \vspace{-0.15in}
\end{table}


\subsection{Empirical Results for Frequency of Spills}
\label{sec:empirical}
\begin{figure}[htb] 
  	\centering
	\includegraphics[width=3.2in]{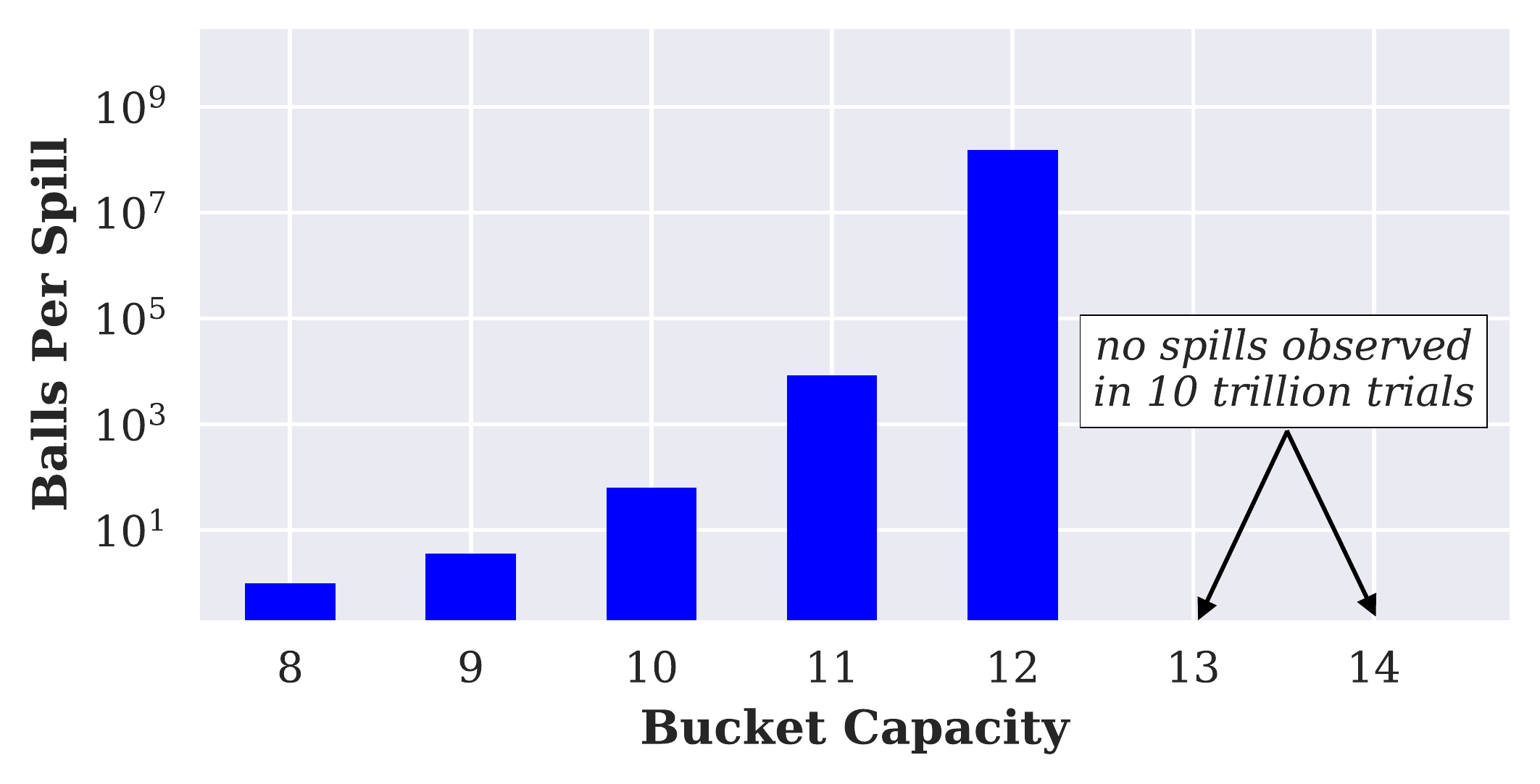}
	\vspace{-0.15in}
	\caption{Frequency of bucket spills, as bucket capacity is varied. As bucket-capacity increases from 8 to 14 (i.e. extra-tags per set increase from 0\% to 75\%), bucket spills (equivalent to SAE) become more infrequent.}
	\label{fig:spills_12ways} 
\end{figure}

Figure~\ref{fig:spills_12ways} shows the average number of balls thrown per bucket spill, analogous to the number of line installs required to cause an SAE on average. As bucket capacity increases from 8 to 14, there is a considerable reduction in the frequency of spills. When the bucket capacity is 8, there is a spill on every throw as each bucket has 8 balls on average. As bucket capacity increases to 9~/~10~/~11~/~12, the spill frequency decreases to once every 4~/~60~/~8000~/~160Mn balls. For bucket capacities of 13 and 14, we observe no bucket spills even after 10~trillion ball throws. These results show that as the number of extra tags increases, the probability of an SAE in Mirage decreases super-exponentially (better than squaring on every extra way). With 12 ways/skew (50\% extra tags), Mirage has an SAE every 160 million installs (equivalent to every 0.16 seconds).

\vspace{0.1in}

While this empirical analysis is useful for estimating the probability of an SAE with up to 12 ways/skew, increasing the ways/skew further makes the frequency of SAE super-exponentially less. Hence, it is impractical to empirically compute the probability of SAE in a reasonable amount of time beyond 12 ways/skew (an experiment with 10 trillion ball throws already takes a few days to simulate). \ignore{Hence, such an empirical analysis cannot provide the SAE probability for a design with 14 ways/skew. }To estimate the probability of SAE for a Mirage design with 14 ways/skew, we develop an analytical model, as described in the next section.  

\begin{table}[ht]
  \begin{center}
    \begin{small}
      \caption{Terminology used in the analytical model  }
        \renewcommand{\arraystretch}{1.3}
    \setlength{\tabcolsep}{4 pt}
\resizebox{3.4in}{!}{    
      \begin{tabular}{|c|l|} \hline

\bf Symbol & \bf Meaning \\ \hline
$\Pr{(n=N)}$ & Probability that a Bucket contains $N$ balls \\
$\Pr{(n\leq N)}$ & Probability that a Bucket contains $\leq N$ balls\\
$\Pr{(X \rightarrow Y)}$  & Probability that a Bucket with $X$ balls transitions to $Y$ balls  \\
$W$ & Capacity of a Bucket (beyond which there is a spill)\\
$B_{tot}$ & Total number of Buckets (32K) \\
$b_{tot}$ & Total number of Balls (256K) \\ \hline
      \end{tabular}
      }
      \label{table:terminology_analytical}
 \end{small}
\end{center}
  \vspace{-0.15 in}
\end{table}

\subsection{Analytical Model for Bucket Spills}

\label{sec:analytical_model}
\begin{figure}[htb] 
  	\centering
	\includegraphics[width=3.2in]{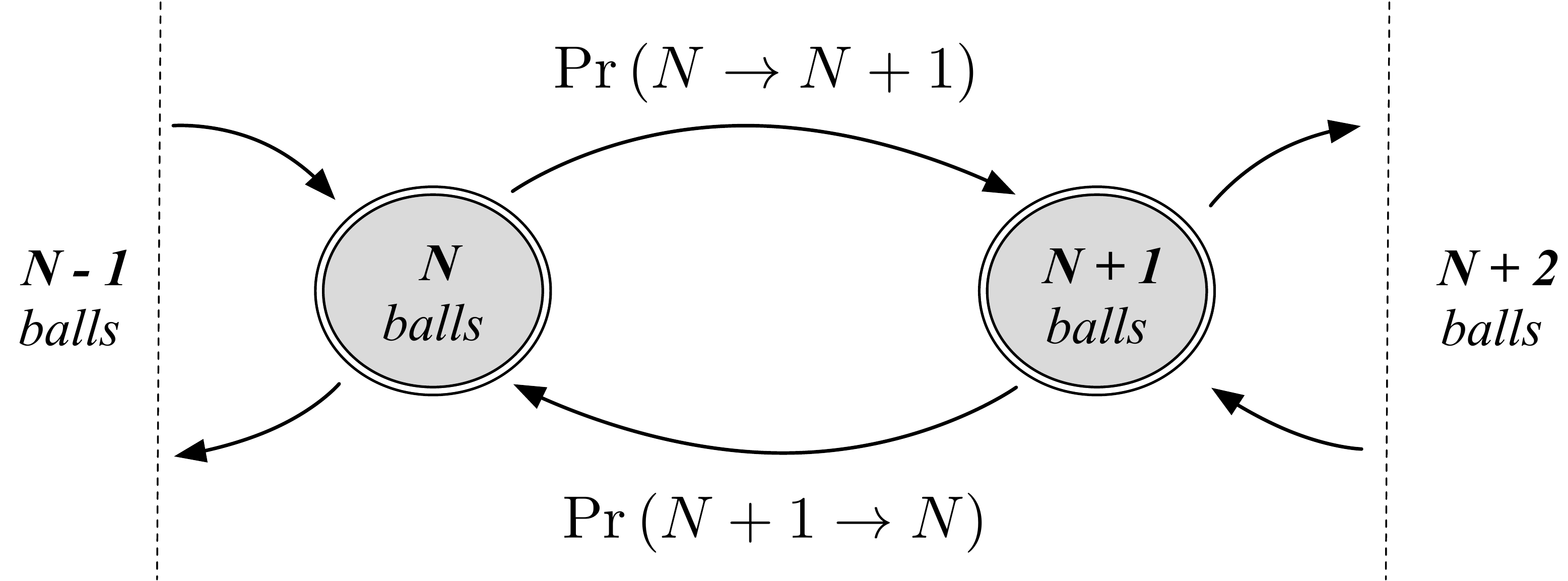}
	\vspace{-0.075in}
	\caption{Bucket state modeled as a Birth-Death chain, a Markov Chain where the state variable \textit{N} (number of balls in a bucket) increases or decreases by one at a time, due to a birth (insertion) or death (deletion) of a ball. }
	\label{fig:birth_death} 
	\vspace{-0.1in}
\end{figure}

To estimate the probability of bucket spills analytically, we start by modeling the behavior of our buckets and balls system in a spill-free scenario (assuming unlimited capacity buckets). We model the bucket-state, i.e.  the number of balls in a bucket, as a Birth-Death chain~\cite{birthdeath}, a type of Markov chain where the state-variable (number of balls in a bucket) only increases or decreases by 1 at a time due to birth or death events (ball insertion or deletions), as shown in Figure~\ref{fig:birth_death}. 

We use a classic result for Birth-Death chains, that in the steady-state\ignore{ (if it exists)\footnote{We can intuitively see why our buckets and balls model of the cache has a steady-state -- (a) the number of balls in a bucket cannot grow unbounded as buckets have finite capacity. (b) experimentally, we observe  the state-probabilities stabilize due to damping behavior of the random ball removal, that penalizes buckets with more balls.}}, the probability of each state converges to a steady value and the net rate of conversion between any two states becomes zero. Applying this result to our model in Figure~\ref{fig:birth_death}, we can equate the probability of a bucket with N balls transitioning to N+1 balls and vice-versa to get Equation~\ref{eq:1}. The terminology used in our model is shown in Table~\ref{table:terminology_analytical}.


\vspace{-0.075in}
\begin{equation}
\label{eq:1}
\Pr{(N \rightarrow  N+1)} = \Pr{(N+1 \rightarrow  N)}
\vspace{0.1in}
\end{equation}
\vspace{-0.15in}

To calculate $\Pr{(N \rightarrow  N+1)}$, we note that a bucket with N balls transitions to N+1 balls on a ball insertion if: (1)~the buckets chosen from both Skew-1 and Skew-2 have N balls; \textit{or} (2)~bucket chosen from Skew-1 has N balls and from Skew-2 has more than N balls; \textit{or} (3)~bucket chosen from Skew-2 has N balls and from Skew-1 has more than N balls. Thus, if the probability of a bucket with N balls is  $\Pr{(n=N)}$, the probability it transitions to N+1 balls is given by Equation~\ref{eq:2}.

{
\medmuskip=1mu
\thinmuskip=1mu
\thickmuskip=1mu
\begin{equation}
\label{eq:2}
  \Pr{(N \rightarrow N+1)}  = \Pr{(n=N)}^2 + 2*\Pr{(n=N)}*\Pr{(n>N)}
\end{equation}
}

\ignore{
{
\medmuskip=1mu
\thinmuskip=1mu
\thickmuskip=1mu
\begin{align}
\begin{split}\label{eq:2}
  \Pr{(N \rightarrow N+1)}  ={}& \Pr{(n=N)}^2 + 2*\Pr{(n=N)}*\Pr{(n>N)}
\end{split}\\[10pt]
\begin{split}\label{eq:3}
                            ={}& \Pr{(n=N)}^2 + 2*\Pr{(n=N)} \\                                           
                            &\phantom{\Pr{(n=N)}^2}- 2*\Pr{(n=N)}*\Pr{(n\leq N)} 
\end{split}
\end{align}
}
}
To calculate $\Pr{(N+1 \rightarrow  N )}$, we note that a bucket with N+1 balls transitions to N balls only on a ball removal. As a random ball is selected for removal from all the balls, the probability that a ball in a bucket with $N+1$ balls is selected for removal equals the fraction of balls in such buckets. If the number of buckets equals $B_{tot}$ and the number of balls is $b_{tot}$, the probability of a bucket with $N+1$ balls losing a ball (i.e. the fraction of balls in such buckets), is given by Equation~\ref{eq:3}.


\vspace{-0.025in}
\begin{equation}
\label{eq:3}
\Pr{(N+1 \rightarrow N)} = \frac{\Pr{(n=N+1)}*B_{tot}*(N+1)}{b_{tot}}
\end{equation}

Combining Equation~\ref{eq:1},~\ref{eq:2},~and~\ref{eq:3}, and placing $B_{tot}/b_{tot} = 1/8$, (the number of buckets/balls) we get the probability of a bucket with N+1 balls\ignore{ as a function of the probabilities of buckets with N balls or less}, as given by Equations~\ref{eq:4} and \ref{eq:5}.

\vspace{-0.05in}
{
\medmuskip=1mu
\thinmuskip=1mu
\thickmuskip=1mu

\begin{align}
\begin{split}\label{eq:4}
 \Pr{(n=N+1)} = \frac{8}{N+1}* {}& \bigg(\Pr{(n=N)}^2  \\     &  + 2*\Pr{(n=N)}*\Pr{(n>N)}\bigg)
\end{split}\\
\begin{split}\label{eq:5}
 = \frac{8}{N+1}* {}& \bigg(\Pr{(n=N)}^2 + 2*\Pr{(n=N)}\\
                                &- 2*\Pr{(n=N)}*\Pr{(n\leq N)}\bigg)
\end{split}
\end{align}
}

\ignore{
{
\medmuskip=2mu
\thinmuskip=2mu
\thickmuskip=2mu
\begin{equation}
\begin{aligned}
\label{eq:5}
\Pr{(n=N+1)} = \frac{8}{N+1}* &\bigg(\Pr{(n=N)}^2 + 2*\Pr{(n=N)}\\
                                &- 2*\Pr{(n=N)}*\Pr{(n\leq N)}\bigg)
\end{aligned}
\end{equation}
}
}


As $n$ grows, $\Pr{(n=N)} \rightarrow 0$ and  $\Pr{(n > N)} \ll \Pr{(n = N)}$ given our empirical observation that these probabilities reduce super-exponentially. Using these conditions Equation~\ref{eq:4} can be simplified to Equation~\ref{eq:6} for larger $n$.


\vspace{-0.05in}
\begin{equation}
\label{eq:6}
\Pr{(n=N+1)} = \frac{8}{N+1}*\Pr{(n=N)}^2 
\end{equation}

\begin{figure}[htb]
  	\centering
	\includegraphics[width=3.2in]{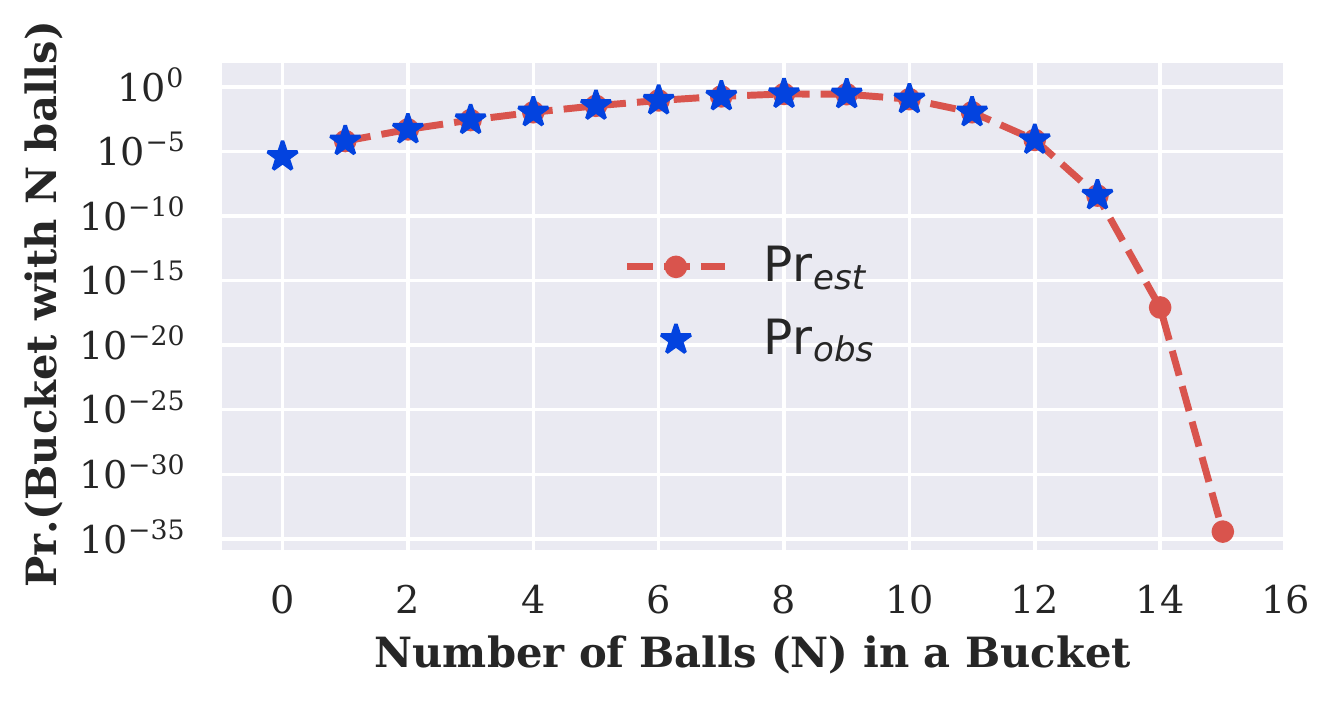}
	\vspace{-0.1in}
	\caption{Probability of a Bucket having N balls -- Estimated analytically ($\Pr_{est}$) and Observed ($\Pr_{obs}$)}
	\label{fig:bucket_prob} 
	\vspace{-0.05in}
\end{figure}

From our simulation of 10 trillion balls, we obtain probability of a bucket with no balls as $\Pr_{obs}{(n=0)}=4\times10^{-6}$. Using this value in Equation~\ref{eq:5}, we recursively calculate $\Pr_{est}(n=N+1)$ for $N\in[1,10]$ and then use Equation~\ref{eq:6} for $N\in[11,14]$, when the probabilities become less than 0.01. Figure~\ref{fig:bucket_prob} shows the  empirically observed ($\Pr_{obs}$) and analytically estimated ($\Pr_{est}$) probability of a bucket having N balls. $\Pr_{est}$ matches $\Pr_{obs}$ for all available data-points.

Figure~\ref{fig:bucket_prob} shows that the probability of a set having N lines decreases double-exponentially beyond 8 lines per set (the average number of data-lines per set). For N~=~13~/~14~/~15, the probability reaches $10^{-9}$~/~$10^{-17}$~/~$10^{-35}$. This behavior is due to two reasons -- (a) for a set to get to N+1 lines, a new line must map to two sets with at least N lines; (b) a set with a higher number of lines is more likely lose a line due to random global eviction. Using these probabilities, we estimate the frequency of SAE in the next section.




\subsection{Analytical Results for Frequency of Spills}
\ignore{We first calculate the probability of a bucket-spill without relocation. }For a bucket of capacity \textit{W}, the spill-probability (without relocation) is the probability that a bucket with $W$ balls gets to $W+1$ balls. By setting $N=W$ in Equation~\ref{eq:2} and $\Pr{(n>W)}=0$, we get the spill-probability as Equation~\ref{eq:7}.

\begin{equation}\label{eq:7}
 \text{Pr} _{spill}{} = \Pr{(W \rightarrow W+1)}  =\Pr{(n=W)}^2
  \end{equation}

\begin{figure}[htb]
  	\centering
	\includegraphics[width=3.2in]{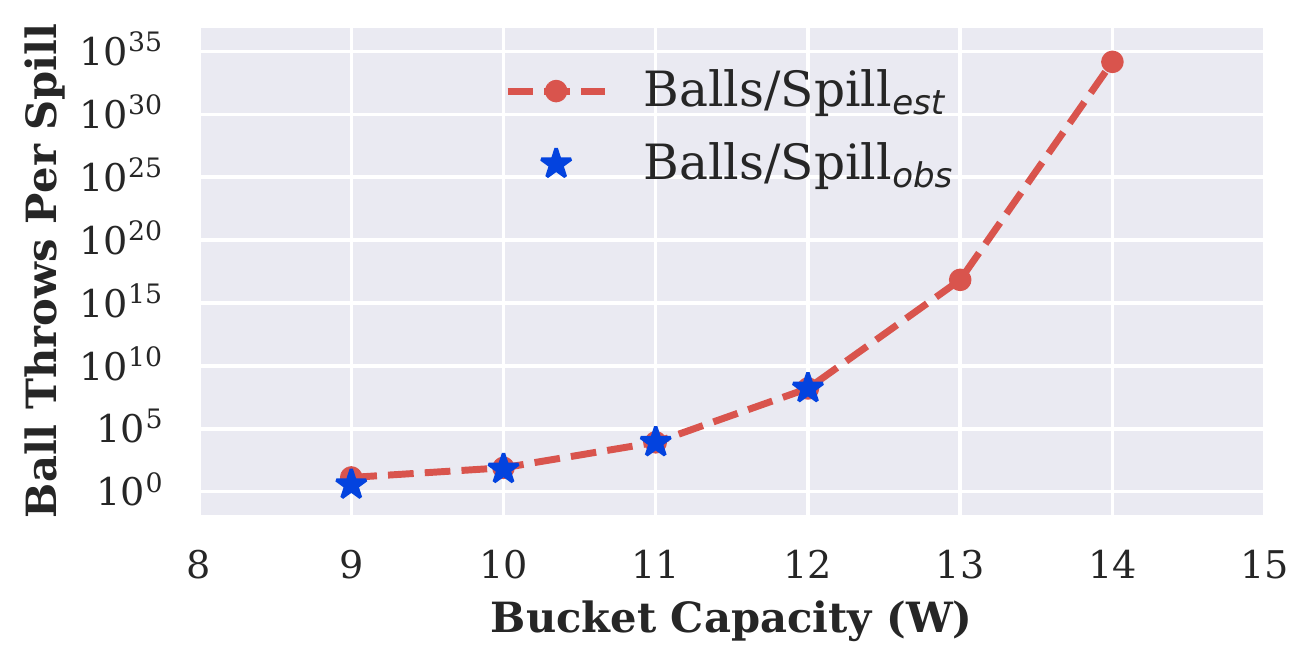}
	\vspace{-0.15in}
	\caption{Frequency of bucket-spill, as bucket-capacity varies -- both analytically estimated (Balls/Spill$_{est}$) and empirically observed (Balls/Spill$_{obs}$) results are shown.}
	\label{fig:bucket_spills} 
	\vspace{-0.15in}
\end{figure}


Figure~\ref{fig:bucket_spills} shows the frequency of bucket-spills (SAE) estimated by using $\Pr_{est}{(n=W)}$, from Figure~\ref{fig:bucket_prob}, in Equation~\ref{eq:7}. The estimated values (Balls/Spill$_{est}$) closely match the empirically observed values (Balls/Spill$_{obs}$) from Section~\ref{sec:empirical}. As the number of tags per set, i.e. bucket-capacity ($W$) increases, the rate of SAE, i.e. the frequency of bucket-spills shows a double-exponential reduction (which means the exponent itself is increasing exponentially). The probability of a spill with $x$ extra ways is of the form $P^{(2^x)}$; therefore with 5-6 extra ways, we get an extremely small probability of spill as the exponent term reaches 32 -- 64. For $W$~=~12~/~13~/~14, an SAE occurs every $10^{8}$~/~$10^{16}$~/~$10^{34}$ line installs.  Thus, the default Mirage design with 14-ways per set, with a rate of one SAE in $10^{34}$ line installs (i.e. once in $10^{17}$ years), effectively provides the security of a fully associative cache.

\section{Protecting against Shared-Memory Attacks}
\label{sec:FlushReload}

Thus far, we have focused primarily on attacks that cause eviction via set conflicts and without any shared data between the victim and the attacker.  If there is shared-memory between the victim and the attacker, attacks such as Flush +Reload~\cite{yaromFlushReload}, Flush+Flush~\cite{FlushFlush}, Invalidate+Transfer~\cite{InvalidateTransfer}, Flush+Prefetch~\cite{FlushPrefetch}, Thrash+Reload~\cite{Netspectre}, Evict+Reload~\cite{EvictReload}, etc. are possible, where an attacker evicts the shared line from the cache using {\em clflush} instruction or cache-thrashing~\cite{Netspectre} or by accessing the line's eviction-set~\cite{EvictReload}, and issues subsequent loads or flushes~\cite{FlushFlush} to the line while measuring its latency to monitor victim accesses to that line. 
We describe how Mirage is protected against these attacks based on the type of shared memory being attacked. 

\vspace{0.05in}
\noindent\textbf{Shared Read-only Memory:}  Attacks on shared read-only memory addresses are prevented in Mirage by placing distrusting programs (victim and attacker) in different security domains and maintaining duplicate copies of shared lines in the cache for each security domain. Such duplication ensures that a load on a shared-address from one domain does not hit on the copy of another domain (similarly flush from one domain does not evict another's copy) and has been used in several prior secure cache works~\cite{DAWG, ScatterCache,Hybcache}. For example, Scatter-Cache (SCv1)~\cite{ScatterCache} uses Security-Domain-ID (SDID) concatenated with the physical line-address as input to the set index derivation function (IDF),
allowing a shared address to map to different sets for different domains and get duplicated. Mirage uses an IDF construction identical to Scatter-Cache SCv1 
and similarly duplicates shared lines across domains.


However, we observe that relying on the IDF to create duplicate copies has a weakness: it can allow a shared-memory address in two different SDIDs to map to the same set in a skew with a small probability ($1/number\text{-}of\text{-}sets$), which can result in a single copy of the line. To guarantee duplicate copies of a line across domains even in this scenario, Mirage stores the SDID of the domain installing the line along with the tag of the line, so that a load (or a flush) of a domain hits on (or evicts) a cache line only if the SDID matches along with the tag-match. Mirage stores 8-bit SDID supporting up to 256 security domains (similar to DAWG~\cite{DAWG}), which adds $<$3\% LLC storage overhead; however more or fewer SDID can be supported without any limitations in Mirage. 




\vspace{0.05in}
\noindent\textbf{Shared Writable Memory:} It is infeasible to duplicate shared writeable memory across domains, as such a design is incompatible with cache-coherence protocols~\cite{ScatterCache,DAWG}. To avoid attacks on such memory, we require that writable shared-memory is not used for any sensitive computations and only used for data-transfers incapable of leaking information.
\section{Discussion}


\subsection{Requirements on Randomizing Function} 

The randomizing function used to map addresses to cache sets in each skew is critical in ensuring balanced availability of invalid tags across sets and eliminating SAE. We use a cryptographic function (computed with a secret key in hardware), so that an adversary cannot arbitrarily target specific sets.  This is also  robust to \textit{shortcut attacks}~\cite{SP21:SystematicRandCaches}, which can exploit vulnerabilities in the algorithm to deterministically engineer collisions. Furthermore, the random-mapping for each skew must be mutually independent to ensure effective load-balancing and minimize naturally occurring collisions, as required by power-of-2-choices hashing~\cite{mitzenmacherThesis}. We satisfy both requirements using a cryptographic hash function constructed using the PRINCE cipher, using separate keys for each skew. Other ciphers and cryptographic hashes that satisfy these requirements may also be used to implement Mirage. 



\subsection{Key Management in Mirage}
The secret keys used in Mirage for the randomizing set-index derivation function are stored in hardware and not visible to any software including the OS. As no information about the mapping function leaks in the absence of SAE in Mirage, by default Mirage does not require continuous key-refreshes like CEASER~/~CEASER-S  \cite{micro18:CEASER,isca19:SkewedCEASER} or keys to be provisioned per domain like  Scatter-Cache\cite{ScatterCache}). 
We recommend that the keys used in Mirage be generated at boot-time within the cache controller (using a hardware-based secure pseudorandom number generator), with the capability to refresh the keys in the event of any key or mapping leakage.
For example, all prior randomized cache designs become vulnerable to conflict-based attacks if the adversary guesses the key via brute-force (1 in $2^{64}$ chance) or if the mappings leak via attacks unknown at the time of designing the defense, as they have no means of detecting such a breakdown in security. On the other hand, Mirage has the capability to automatically detect a breach in security via even hypothetical future attacks, as any subsequent conflict-based attack requires the orchestration of SAE, which do not occur in Mirage under normal operation. If multiple SAE are encountered indicating that the mapping is no longer secret, Mirage can adapt by transparently refreshing its keys (followed by a cache flush) to ensure continued security.

\subsection{Security for Sliced LLC Designs}
\ignore{Thus far, we have assumed an LLC that is implemented as a monolithic structure with a single tag-store. }
Recent Intel CPUs have LLCs that consist of multiple smaller physical entities called slices (each a few MBs in size), with separate tag-store and data-store structures for each slice. In such designs, Mirage can be implemented at the granularity of a slice (with per-slice keys) and can guarantee global evictions within each slice. We analyzed the rate of SAE for an implementation of Mirage per 2MB slice (2048 sets, as used in Intel CPUs) with the tag-store per slice having 2 skews and 14-ways per skew and observed it to be one SAE in $2\times10^{17}$ years, whereas a monolithic 16MB Mirage provides a rate of once in $5\times10^{17}$ years. Thus, both designs (monolithic and per-slice) provide protection for a similar order of magnitude (and well beyond the system lifetime).



\subsection{Security as Baseline Associativity Varies}
The rate of SAE strongly depends on the number of ways provisioned in the tag-store.  Table~\ref{table:security_assoc} shows the rate of SAE for a 16MB LLC, as the baseline associativity varies from 8 ways -- 32 ways. As the baseline associativity varies, with just 1 extra way per skew, the different configurations have an SAE every 13 -- 14 installs.  However, adding each extra way squares the rate successively as per Equation~\ref{eq:7}. Following the double-exponential curve of Figure~\ref{fig:bucket_spills}, the rate of an SAE goes beyond once in $10^{12}$ years (well beyond system lifetime) for all three configurations within 5--6 extra ways.


\begin{table}[h]
  \begin{center}
 \vspace{-0.1 in}
  \caption{Cacheline installs Per SAE in Mirage as the baseline associativity of the LLC tag-store varies}
   \renewcommand{\arraystretch}{1.3}
    \setlength{\tabcolsep}{4pt}
    \resizebox{3.4in}{!}{    
      \begin{small}
      \begin{tabular}{|c||c|c|c|} \hline
      
        \textbf{LLC Associativity}      & \textbf{8-ways} &\textbf{ 16-ways (default)} & \textbf{32-ways}  \\ \hline \hline
        \textbf{1 extra way/skew } & 13 ($<$ 20ns) & 14 ($<$ 20ns) & 14 ($<$ 20ns) \\ \hline
        \textbf{5 extra ways/skew} & $10^{21}$ ($10^{4}$ yrs) & $10^{16}$ (2 yrs) & $10^{14}$ (3 days) \\ \hline
        \textbf{6 extra ways/skew} & $10^{43}$ ($10^{26}$ yrs) & $10^{34}$ ($10^{17}$ yrs) & $10^{29}$ ($10^{12}$ yrs) \\ \hline
      \end{tabular}
 
      \end{small}
      }
\label{table:security_assoc}
\end{center}
\vspace{-0.3 in}
\end{table}

\subsection{Implications for Other Cache Attacks}
\textbf{Replacement Policy Attacks:} Reload+Refresh~\cite{ReloadRefresh} attack exploited the LLC replacement policy to influence eviction-decisions within a set, and enable a side-channel stealthier than Prime+Probe or Flush+Reload. Mirage guarantees global evictions with random replacement, that has no access-dependent state. This ensures that an adversary cannot influence the replacement decisions via its accesses, making Mirage immune to any such replacement policy attacks.

\vspace{0.1in}

\noindent\textbf{Cache-Occupancy Attacks:} Mirage prevents an adversary that observes an eviction from gaining any information about the address of an installed line. However, the fact that an eviction occurred continues to be observable, similar to prior works such as Scatter-Cache~\cite{ScatterCache} and HybCache~\cite{Hybcache}. Consequently, Mirage and these prior works, are vulnerable to attacks that monitor the cache-occupancy of a victim by measuring the number of evictions, like a recent attack~\cite{CacheWebsiteFingerprinting} that used cache-occupancy as a signature for website-fingerprinting. The only known way to effectively mitigate such attacks is static partitioning of the cache space. In fact, Mirage can potentially provide a substrate for global partitioning of the data-store that is more efficient than the current way/set partitioning solutions to mitigate such attacks. We leave the study extending Mirage to support global partitions for future work.

\section{Mirage with Cuckoo-Relocation} 
\ignore{The reduction in episodes of SAE with Mirage depends on the number of extra tags provisioned, as shown in Table~\ref{table:overflows_choice_indexing}. }

The default design for Mirage consists of 6 extra ways / skew (75\% extra tags) that avoids SAE for well beyond the system lifetime. If Mirage is implemented with fewer extra tags (e.g. 4 extra ways/skew or 50\% extra tags), it can encounter SAE as frequently as once in 0.16 seconds. To avoid an SAE even if only 50\% extra tags are provisioned in Mirage, we propose an extension of Mirage that relocates conflicting lines to alternative sets in the other skew, much like Cuckoo Hashing~\cite{cuckoo_hashing}. We call this extension {\em Cuckoo-Relocation}.

\subsection{Design of Cuckoo-Relocation}
\begin{figure}[htb] 
  	\centering
		\includegraphics[width=3.4in]{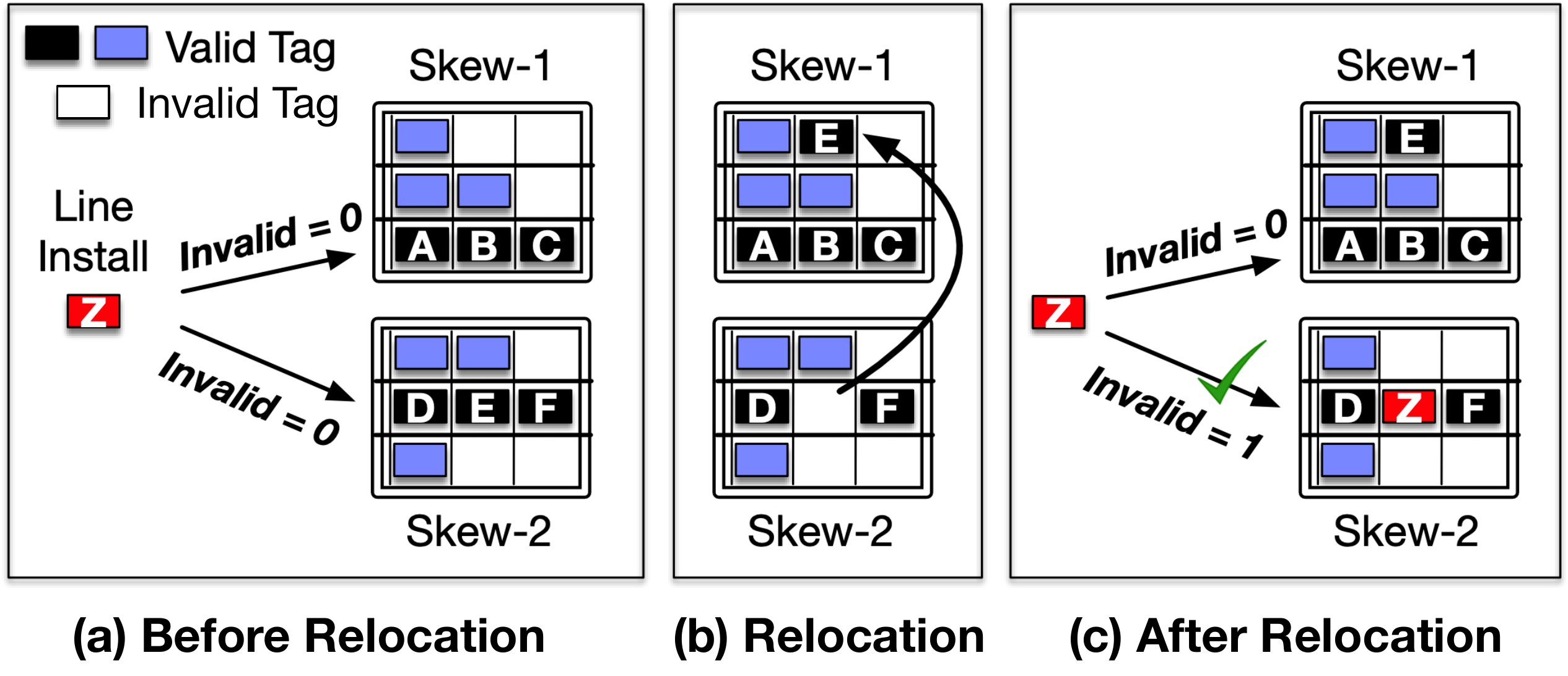}
      \vspace{-0.25 in}
      \caption{ Cuckoo Relocation, a technique to avoid an SAE if Mirage is implemented with 50\% extra tags.} 
      \vspace{-0.1in}
	\label{fig:cuckoo_relocation} 
	
\end{figure}

We explain the design of Cuckoo-Relocation using an example shown in Figure~\ref{fig:cuckoo_relocation}. An SAE is required when an incoming line (Line Z) gets mapped in both skews to sets that have no invalid tags (Figure~\ref{fig:cuckoo_relocation}(a)). To avoid an SAE, we need an invalid tag in either of these sets. To create such an invalid tag, we randomly select a candidate line (Figure~\ref{fig:cuckoo_relocation}(b)) from either of these sets and relocate it to its alternative location in the other skew. If this candidate maps to a set with an invalid tag in the other skew, the relocation leaves behind an invalid tag in the original set, in which the line to be installed can be accommodated without an SAE, as shown in Figure~\ref{fig:cuckoo_relocation}(c). If the relocation fails as the alternative set is full, it can be attempted again with successive candidates till a certain number of maximum tries, after which an SAE is incurred.  For Mirage with 50\% extra tags, an SAE is infrequent even without relocation (less than once in 100 million installs). So in the scenario where an SAE is required, it is likely that other sets have invalid tags and relocation succeeds.

\subsection{Results: Impact of Relocation on SAE}
For Mirage with 50\% extra tags, the chance that a relocation fails is approximately $p = 1/\text{sets per skew}$. This is because, at the time of an SAE (happens once in 100 million installs), it is likely that the only full sets are the ones that are currently indexed (i.e. only 1 set per skew is full). For relocation to fail for a candidate, the chance that its alternative set is full is hence approximately $p = 1/\text{sets per skew}$. After $n$ relocation attempts, the chance that all relocation attempts fail and an SAE is incurred, is approximately $p^n$.

 Table~\ref{table:sec_time_relocation} shows the rate of SAE for Mirage with 50\% extra tags and Cuckoo-Relocation\ignore{ (we call this {\em Mirage-Lite})}, as the maximum number of relocation attempts is varied. Attempting relocation for up to 3 lines is sufficient to ensure that an SAE does not occur in system-lifetime (SAE occurs once in 22000 years). We note that attempting relocation for up to 3 lines can be done in the shadow of a memory access on a cache-miss.

\begin{table}[h]
  \begin{center}
\vspace{-0.1 in}
   \renewcommand{\arraystretch}{1.3}
    \setlength{\tabcolsep}{3pt}
    \caption{Frequency of SAE in Mirage with 50\% extra tags (4 extra ways/skew) as number of relocation attempts increase}
    \resizebox{3.4in}{!}{    
      \begin{small}
      \begin{tabular}{|c|c|c|c|c|} \hline
      
        \textbf{Max Relocations}      & 0 & 1 & 2 & 3  \\ \hline \hline
        \textbf{Installs per SAE}                      & $2 \times 10^8$  & $3 \times 10^{12}$  & $4 \times 10^{16}$  & $7 \times 10^{20}$  \\ \hline
        \textbf{Time per SAE }               &   0.16 seconds &	45 minutes &	1.3 years &	22,000 years \\ \hline
      \end{tabular}
      \end{small}
      }
\label{table:sec_time_relocation}
\end{center}
\vspace{-0.25 in}
\end{table}

\subsection{Security Implications of Relocation}
For Mirage with 50\% extra tags, up to 3 cuckoo relocation are done in the shadow of memory access on a cache-miss. A typical adversary, capable of only monitoring load latency or execution time, gains no information about when or where relocations occur as -- (1)~Relocations do not stall the processor or cause memory traffic, they only rearrange cache entries within the tag-store; (2)~A relocation occurs infrequently (once in 100 million installs) and any resultant change in occupancy of a set has a negligible effect on the probability of an SAE. If a future adversary develops the ability to precisely monitor cache queues and learn when a relocation occurs to perceive a potential conflict,\ignore{  If such modes of vulnerability become a concern,} we recommend implementing Mirage with a sufficient extra tags (e.g. 75\% extra tags) such that no relocations are needed in the system lifetime. 


\vspace{-0.05in}
\section{Performance Analysis}

In this section, we analyze the impact of Mirage on cache misses and system performance.  As relocations are uncommon, we observe that performance is virtually identical for both with and without relocations.  So, we discuss the key results only for the default Mirage design (75\% extra tags).


\subsection{Methodology}

Similar to prior works on randomized caches~\cite{ScatterCache,micro18:CEASER,isca19:SkewedCEASER,PhantomCache}, we use a micro-architecture simulator to evaluate performance. We use an in-house simulator that models an inclusive 3-level cache hierarchy (with private L1/L2 caches and shared L3 cache) and DRAM in detail, and has in-order x86 cores supporting a subset of the instruction-set. The simulator input is a 1 billion instructions long program execution-trace (consisting of instructions and memory-addresses), chosen from a representative phase of a program using the Simpoints sampling methodology~\cite{simpoints} and obtained using an Intel Pintool~\cite{Pin}. We validated the results of our simulator with RISC-V RTL  (Appendix~\ref{sec:riscv_validation}) and Gem5 (Appendix~\ref{sec:gem5_validation}) simulations. 

As our baseline, we use a non-secure 16-way, 16MB set-associative LLC configured as shown in Table~\ref{table:config}. For Mirage, we estimate the LLC access latency using RTL-synthesis of the cache-lookup circuit (Section~\ref{sec:access_lat_synthesis}) and Cacti-6.0~\cite{cacti6.0} (a tool that reports timing, area, and power for caches), and show that it requires 4 extra cycles compared to the baseline (3-cycles for PRINCE cipher and 1 extra cycle for tag and data lookup). For comparisons with the prior state-of-the-art, we implement Scatter-Cache with 2-skews, 8 ways/skew and use PRINCE cipher for the hash function for set-index derivation, that adds 3 cycles to lookups compared to baseline (to avoid an unfair advantage to Mirage, as Scatter-Cache~\cite{ScatterCache} originally used a 5-cycle QARMA-64 cipher). We evaluate 58 workloads, including all 29 SPEC CPU2006 benchmarks (each has 8 duplicate copies running on 8 cores) and 29 mixed workloads (each has 8 randomly chosen SPEC benchmarks) All performance averages reported in subsequent sections are averaged over all 58 workloads, unless mentioned otherwise.

\begin{figure*}[htb] 
  	\centering
  \vspace{-0.1 in}
\includegraphics[width=6.5in]{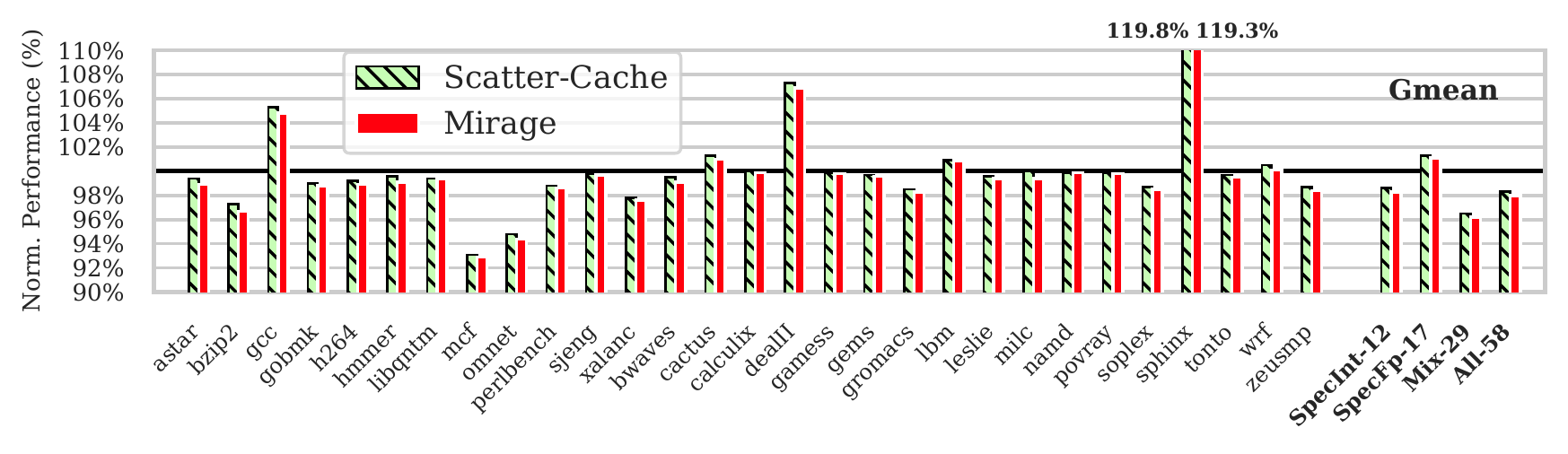}
 \vspace{-0.25in}
\caption{Performance of Mirage and Scatter-Cache normalized to Non-Secure Baseline (using weighted speedup metric). Over 58 workloads, Mirage has a slowdown of 2\%, while Scatter-Cache has a slowdown of 1.7\% compared to the Non-Secure LLC.}
 \vspace{-0.15in}
	\label{fig:perf} 
	
\end{figure*}

\begin{table}[h]
  \begin{center}
    \begin{small}
      \caption{Baseline System Configuration}
   \vspace{0.05in}
        \renewcommand{\arraystretch}{1.5}
     \setlength{\tabcolsep}{3 pt}
      \resizebox{3.4in}{!}{
     
      \begin{tabular}{|l|l|} \hline

\multicolumn{2}{|c|}{\bf Processor and Last-level Cache} \\ \hline
Core      &   8-cores, In-order Execution, 3GHz    \\ 
L1 and L2 Cache Per Core  & L1-32KB, L2-256KB, 8-way, 64B linesize \\ \hline


\multirow{2}{*}{LLC (shared across cores)}  &  16MB, 16-way Set-Associative, 64B linesize \\ 
                        &   LRU Replacement Policy, 24 cycle lookup \\
 \hline
\multicolumn{2}{|c|}{\bf DRAM Memory-System} \\ \hline

Frequency, tCL-tRCD-tRP      &    800 MHz (DDR 1.6 GHz), 9-9-9 ns  \\
DRAM Organization     &        2-channel  (8-Banks each), 2KB Row-Buffer\\ \hline
      \end{tabular}
}
      \label{table:config}
 \end{small}
\end{center}
 \vspace{-0.2 in}
\end{table}

\subsection{\mbox{Synthesis Results for Cache Access Latency}}
\label{sec:access_lat_synthesis}
Compared to the baseline, the cache access in Mirage additionally requires (a) set-index computation using the PRINCE cipher based hash-function, (b) look-up of 8-12 extra ways of the tag-store, and (c) FPTR-based indirection on a hit to access the data.
We synthesized the RTL for the set-index derivation function with a 12-round PRINCE cipher~\cite{PRINCE} based on a public VHDL implementation~\cite{PRINCE_vhdl}, using Synopsys Design Compiler and FreePDK 15nm 
gate library~\cite{15nmPDK}. A 3-stage pipelined implementation (with 4 cipher rounds/stage) has a delay of 320ps per stage (which is less than a cycle period). Hence, we add 3 cycles to the LLC access latency for Mirage (and Scatter-Cache), compared to the baseline. 

We also synthesized the RTL for FPTR-indirection circuit consisting of AND and OR gates that select the FPTR value of the hitting way among the accessed tags, and a 4-to-16 decoder to select the data-store way using the lower 4-bits of the FPTR (the remaining FPTR-bits form the data-store set-index); the circuit has a maximum delay of 72ps. Using Cactii-6.0~\cite{cacti6.0}, we estimate that lookup of up to 16 extra ways from the tag-store further adds 200ps delay in 32nm technology. To accommodate the indirection and tag lookup delays, we increase the LLC-access latency for Mirage further by 1 cycle (333ps). Overall, Mirage incurs 4 extra cycles for cache-accesses compared to the baseline.
Note that the RPTR-lookup and the logic for skew-selection (counting valid bits in the indexed set for each skew and comparing) require simple circuitry with a delay less than 1 cycle. These operations are only required on a cache-miss and performed in the background while the DRAM-access completes.


\begin{table}[h]
  \begin{center}
 \vspace{-0.1 in}

   \renewcommand{\arraystretch}{1.1}
    \caption{Average LLC MPKI of Mirage and Scatter-Cache}
   \vspace{0.1in}
    \begin{small}

 \resizebox{2.8in}{!}{
      \begin{tabular}{|c||c|c|c|} \hline 
        \textbf{Workloads}  & \textbf{Baseline} & \textbf{Mirage} & \textbf{Scatter-Cache} \\ \hline \hline
        SpecInt-12  & 10.79 & 11.23 & 11.23 \\ \hline
        SpecFp-17   & 8.82 & 8.51  & 8.51  \\ \hline
        Mix-29      & 9.52 & 9.97  & 9.97  \\ \hline \hline
        All-58      & 9.58 & 9.80  & 9.80  \\ \hline
    \end{tabular}
}
\end{small}

\label{table:mpki}
\end{center}
\vspace{-0.2 in}
\end{table}

\vspace{-0.1in}
\subsection{Impact on Cache Misses}
Table~\ref{table:mpki} shows LLC Misses Per 1000 Instructions (MPKI) for the non-secure Baseline,  Mirage, and Scatter-Cache averaged for each workload suite. We observe that all LLC-misses in Mirage in all workloads result in Global Evictions (no SAE), in line with our security analysis.\footnote{Workloads typically do not always access random addresses. But the randomized cache-set mapping used in Mirage ensures accesses always map to random cache-sets, which allows the load-balancing skew-selection to maintain the availability of invalid tags across sets and prevent any SAE.} Compared to the Baseline, Mirage incurs 2.4\% more misses on average (0.2 MPKI extra) as the globally-random evictions from the data-store lack the intelligence of the baseline LRU policy that preserves addresses likely to be re-used. The miss count for Scatter-Cache is similar to Mirage as it uses randomized set-indexing that causes randomized evictions with similar performance implications (however note that all its evictions are SAE that leak information). We observe that randomization can increase or decrease conflict misses for different workloads: e.g., Mirage and Scatter-Cache increase misses by 7\% for \textit{mcf} and \textit{xalanc} while reducing them by 30\% for \textit{sphinx} compared to baseline.

\subsection{Impact on Performance}

Figure~\ref{fig:perf} shows the relative performance for Mirage and Scatter-Cache normalized to the non-secure baseline (based on the \textit{weighted speedup}\footnote{
$Weighted\text{-}Speedup=\sum_{i=0}^{N-1}{IPC\text{-}MC_i/IPC\text{-}SC_i}$  is a popular throughput metric for fair evaluation of $N$-program workloads~\cite{WeightedSpeedup}, where \textit{IPC} stands for Instructions per Cycle, $IPC\text{-}MC_i$ is the IPC of a program-$i$ in multi-program setting, and $IPC\text{-}SC_i$ is the IPC of program-$i$ running alone on the system.
Using \textit{Raw-IPC} as the throughput metric, the slowdown decreases by 0.2\%.
}
metric). 
On average, Mirage incurs a 2\% slowdown due to two factors: increased LLC misses and a 4 cycle higher LLC access latency compared to the baseline. For workloads such as \textit{mcf} or \textit{omnet}, Mirage increases both the LLC misses and access latency compared to a non-secure LLC and hence causes 6\% to 7\% slowdown. On the other hand, for workloads such as \textit{sphinx}, \textit{dealII} and \textit{gcc}, Mirage reduces LLC-misses and improves performance by 5\% to 19\%. In comparison, Scatter-Cache has a lower slowdown of 1.7\% on average despite having similar cache-misses, as it requires 1 cycle less than Mirage for cache accesses (while both incur the cipher latency for set-index calculation, Mirage requires an extra cycle for additional tag-lookups and indirection). 
\vspace{-0.05in}
\subsection{Sensitivity to Cache Size}
\label{sec:cache_sz_sensitivity}
\begin{figure}[htb]
  	\centering
	\includegraphics[width=3.3in]{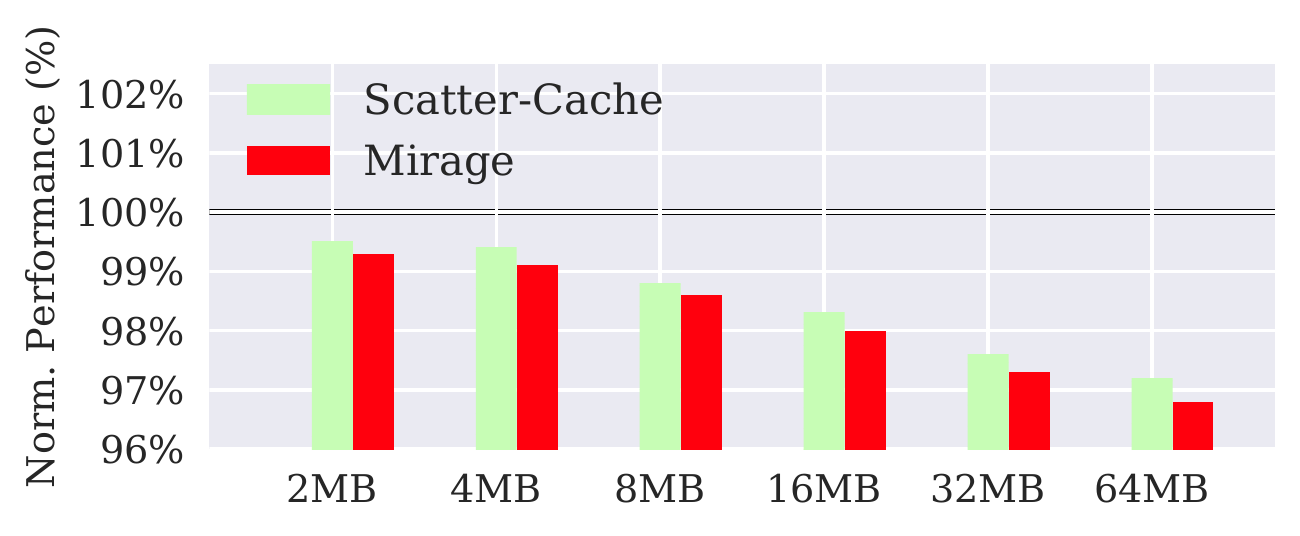}
	\vspace{-0.15in}
	\caption{Sensitivity of Performance to Cache-Size.}
	\label{fig:sensitivity_cachesz} 
	\vspace{-0.05in} 
\end{figure}

Figure~\ref{fig:sensitivity_cachesz} shows the performance of Mirage and Scatter-Cache for LLC sizes of 2MB to 64MB, each normalized to a non-secure design of the same size. As cache size increases, the slowdown for Mirage increases from 0.7\% for a 2MB cache to 3.2\% for a 64MB cache. This is because larger caches have a higher fraction of faster cache-hits that causes the increase in access-latency to have a higher performance impact. Similarly, the slowdown for Scatter-Cache increases from 0.5\% to 2.8\% and is always within 0.4\% of Mirage.






\subsection{Sensitivity to Cipher Latency}
\begin{figure}[htb]
    \vspace{-0.05in}
  	\centering
	\includegraphics[width=3.3in]{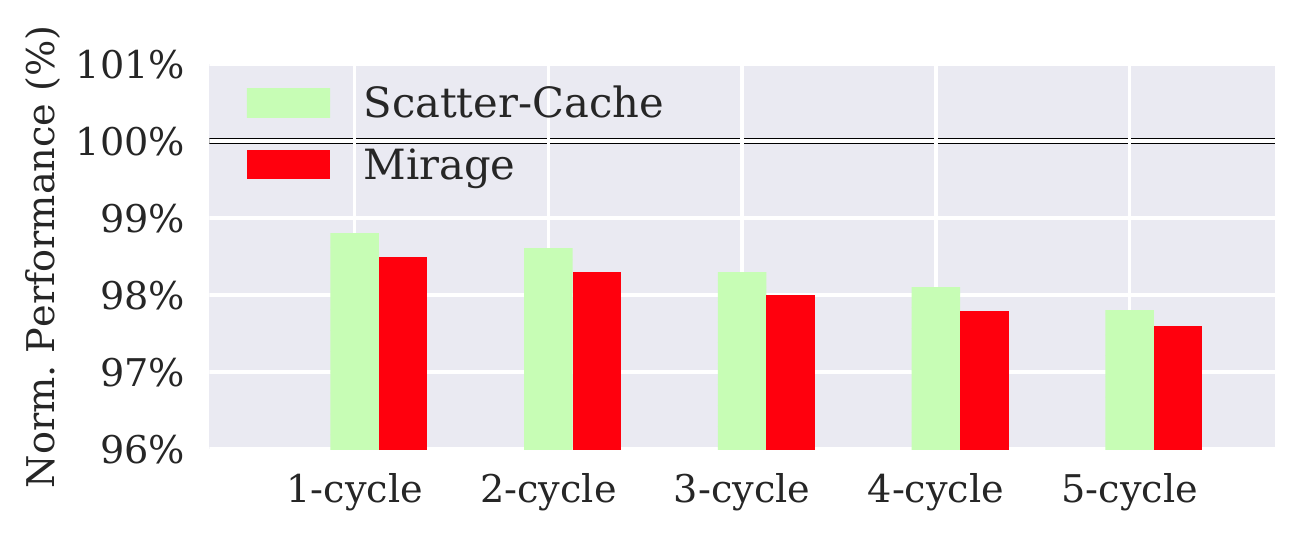}
	\vspace{-0.15in}
	\caption{
	Sensitivity of Performance to Cipher Latency.}
	\label{fig:sensitivity_encrlat} 
\end{figure}

Figure~\ref{fig:sensitivity_encrlat} shows the performance of Mirage and Scatter-Cache normalized to a non-secure baseline LLC, as the latency of the cipher (used to compute the randomized hash of addresses) varies from 1 to 5 cycles. By default, Mirage and Scatter-Cache evaluations in this paper use a 3-cycle  PRINCE-cipher~\cite{PRINCE} (as described in Section~\ref{sec:access_lat_synthesis}), resulting in  slowdowns of 2\% and 1.7\% respectively. Alternatively, a cipher like QARMA-64~\cite{QARMA64} (that was used in the Scatter-Cache paper and assumed to have 5 cycle latency~\cite{ScatterCache}) can also be used in Mirage; this causes Mirage and Scatter-Cache to have higher slowdowns of 2.4\% and 2.2\%. Similarly, future works may design faster randomizing-functions for set-index calculations in randomized caches; a  1-cycle latency randomizing function can reduce slowdown of Mirage and Scatter-Cache to 1.5\% and 1.2\% respectively. The study of faster randomizing functions for Mirage that also have robust randomization that prevents an adversary from discovering eviction-sets via \textit{shortcut attacks}~\cite{SP21:SystematicRandCaches} is an important direction for future work.

\vspace{-0.1in}
\section{Cost Analysis}
For analyzing the storage and power overheads of Mirage, we distinguish the two versions of our design as, {\em Mirage} (default design with 75\% extra tags) and {\em Mirage-Lite} (with 50\% extra tags and relocation). 

\ignore{
\vspace{-0.1in}
\subsection{Impact on Lookup Latency}
\vspace{-0.05in}
The lookup latency in Mirage is affected by a larger tag-store and an extra mux to select the FPTR of a hitting way. Using Cacti-6.0\cite{cacti6.0} (that reports access time, energy, etc. for different cache organizations), with 32nm technology, we estimate that adding up to 16 extra ways adds 0.2~ns delay and an extra mux adds \textasciitilde0.05ns. These delays can be accounted by increasing the latency \ignore{ in both Mirage (12 extra ways) and Mirage-Lite (8 extra ways)} by 1 cycle compared to Scatter-Cache.
}


\subsection{Storage Overheads}
The storage overheads in Mirage are due to (1) extra tag-entries, and (2) FPTR and RPTR, the pointers between tag/data entries, and (3) tag-bits storing full 40-bit line-address (for 46-bit physical address space) to enable address generation for write-backs. This causes a storage overhead of 20\% for Mirage and 17\% for Mirage-Lite compared to the non-secure baseline, as shown in Table~\ref{table:storage}. These overheads are dependent on cache linesize as the relative size of tag-store compared to the data-store reduces at a larger linesize. While we use 64B linesize, a 128B linesize like IBM's Power9 CPUs~\cite{IBMPower9} would reduce these overheads to 9-10\% and a 256B linesize would reduce these to 4-5\%. 

The storage overhead in Mirage is the main driver behind the area overhead, as the extra storage requires millions of gates, whereas all other extra logic for FPTR/RPTR-indirection, PRINCE cipher, etc., can be implemented in few thousand gates (as shown in Section~\ref{sec:logic_overheads}). Using CACTI-6.0~\cite{cacti6.0}, we estimate that an LLC requiring 20\% extra storage consumes approximately 22\% extra area. In terms of a storage-neutral comparison, Mirage has an average slowdown $<$3.5\% compared to a non-secure LLC with 20\% more capacity.

\begin{table}[htb]
  \begin{center}
 \vspace{-0.1 in}
   \renewcommand{\arraystretch}{1.25}
     \setlength{\tabcolsep}{2 pt}
    \caption{Storage Overheads in Mirage for 64B linesize}
    \vspace{0.075in}
    \begin{small}
      \begin{tabular}{|c|c|c|c|c|} \hline
        \multicolumn{2}{|c|}{\textbf{Cache Size}} & \textbf{Baseline}     & \textbf{Mirage} & \textbf{Mirage-Lite}   \\ 
        \multicolumn{2}{|c|}{16MB}               & Set   &  2 skews x & 2 skews x  \\  
        \multicolumn{2}{|c|}{(16,384 Sets)}      & Associative    &  14 ways/skew  &  12 ways/skew   \\\hline\hline
        \multirow{5}{*}{\parbox{0.75cm}{\center Tag Entry}}   & Tag-Bits    & 26    & 40   & 40 \\ 
                                   & Status(V,D)        & 2       & 2      &   2 \\ 
                                   & FPTR                     & --      & 18       &   18 \\
                                   & SDID                     & --      & 8       &   8 \\\hhline{~====}
                                   & Bits/Entry         & 28      & 68     &   68 \\ 
                                   & Tag Entries        & 262,144   & 458,752  & 393,216\\ \hline\hline

        \multicolumn{2}{|c|}{\textbf{Tag-Store Size}} & \textbf{896 KB}     & \textbf{3808 KB}   & \textbf{3264 KB} \\ \hline\hline
        \multirow{4}{*}{\parbox{0.75cm}{\center Data Entry}} & Data-Bits     & 512 & 512     & 512 \\ 
                                   & RPTR                     & --           &  19  &  19 \\\hhline{~====}
                                   & Bits/Entry         & 512          &  531 &  531 \\ 
                                   & Data Entries       & 262,144       &  262,144 &  262,144 \\ \hline\hline
        \multicolumn{2}{|c|}{\textbf{Data-Store Size}} & \textbf{16,384 KB} & \textbf{16,992 KB}   & \textbf{16,992 KB} \\ \hline\hline
        
        \multicolumn{2}{|c|}{\multirow{2}{*}{\textbf{Total Storage}}} & \textbf{17,280 KB}       & \textbf{20,800 KB}  & \textbf{20,256 KB}\\
        \multicolumn{2}{|c|}{} & \textbf{(100\%)}      & \textbf{(120\%)}   & \textbf{(117\%)} \\
        \hline
      \end{tabular} 
    \end{small}   
\label{table:storage}
\end{center}
\vspace{-0.2 in}
\end{table}

\subsection{Power Consumption Overheads}
The larger tag-store in Mirage has a higher static leakage power when idle and also consumes more energy per read/write access. Table~\ref{table:energy} shows the static and dynamic power consumption for Mirage in 32nm technology estimated using CACTI-6.0~\cite{cacti6.0}, which reports the energy/access and static leakage power consumption for different cache organizations. We observe the LLC power is largely dominated by the static leakage power compared to dynamic power (in line with prior CPU power modeling studies~\cite{IPDPS16_LLCPowerModel}). The static power in Mirage (reported by CACTI) increases by 3.5-4.1W (18\%-21\%) in proportion to the storage overheads, whereas the dynamic power, calculated by multiplying the energy/access (from CACTI) by the total LLC-accesses per second (from our simulations), shows an insignificant increase of 0.02W on average. The increase in LLC power consumption of 4W (21\%) in Mirage is quite small compared to the overall chip power budget, with comparable modern 8-core Intel/AMD CPUs having power budgets of 95-140W~\cite{AnandTech_Power_Consumption}.

\begin{table}[!h]
  \begin{center}
 \vspace{-0.1 in}

   \renewcommand{\arraystretch}{1.12}
     \setlength{\tabcolsep}{2pt}
    \caption{Energy and Power Consumption for Mirage}
    \begin{small}
    \resizebox{\columnwidth}{!}{
      \begin{tabular}{|c||c|c|c|c|} \hline
            \multirow{2}{*}{\textbf{Design}}         &   
              \textbf{Energy /} & \textbf{Dynamic} & \textbf{Static Leakage} & \textbf{Total}\\
              & \textbf{Access (nJ)} &  \textbf{Power (W) }& \textbf{Power  (W)}& \textbf{Power (W)} \\ \hline \hline
\textbf{Baseline}   & 0.61   & 0.06  & 19.2  & 19.3  \\ \hline 
\textbf{Mirage}     & 0.78   & 0.08  & 23.3  & 23.4  \\ \hline 
\textbf{Mirage-Lite}& 0.73   & 0.08  & 22.7  & 22.8  \\ \hline 
      \end{tabular}
      }
    \end{small}   

\label{table:energy}
\end{center}
\vspace{-0.25 in}
\end{table}

\subsection{Logic Overheads}
\label{sec:logic_overheads}
Mirage requires extra logic for the set-index computation using the randomizing hash-function and FPTR-indirection on cache-lookups, and for load-aware skew-selection and RPTR-indirection based tag-invalidation on a cache-miss. Our synthesis results in 15nm technology show that the PRINCE-based randomizing hash-function occupies 5460 $\text{um}^2$ area or 27766 Gate-Equivalents (GE - number of equivalent 2-input NAND gates) and the FPTR-indirection based lookup circuit requires 132 $\text{um}^2$ area or 670 GE. The load-aware skew-selection circuit (counting 1s among valid bits of 14 tags from the indexed set in each skew, followed by a 4-bit comparison) requires 60  $\text{um}^2$ or 307 GE, while the RPTR-lookup circuit complexity is similar to the FPTR-lookup. Overall, all of the extra logic (including the extra control state-machine) can fit in less than 35,000 GE, occupying a negligible area compared to the several million gates required for the LLC.



\section{Related Work}

Cache design for reducing conflicts (for performance or security) has been an active area of research.  In this section, we compare and contrast Mirage with closely related proposals.

\subsection{\mbox{Secure Caches with High Associativity}}
\label{sec:SecureCacheFullAssoc}
The concept of cache location randomization for guarding against cache attacks was pioneered almost a decade ago, with {\bf RPCache}~\cite{RPCache} and {\bf NewCache}~\cite{NewCache}, for protecting L1 caches. Conceptually, such designs have an indirection-table that is consulted on each cache-access, that allows mapping an address to any cache location. While such designs can be implemented for L1-Caches, there are practical challenges when they are extended to large shared LLCs. For instance, the indirection-tables themselves need to be protected from conflicts if they are shared among different processes. While RPCache prevents this by maintaining per-process tables for the L1 cache, such an approach does not scale to the LLC that may be used by several hundred processes at a time. NewCache avoids conflicts among table-entries by using a Content-Addressable-Memory (CAM) to enable a fully-associative design for the table. However, such a design is not practical for LLCs, which have tens of thousands of lines, as it would impose impractically high power overheads. While Mirage also relies on indirection for randomization, it eliminates conflicts algorithmically using load-balancing techniques, rather than relying on per-process isolation that requires OS-intervention, or impractical fully-associative lookups and CAMs. 

{\bf Phantom-Cache}~\cite{PhantomCache} is a recent design that installs an incoming line in 1 of 8 randomly chosen sets in the cache, each with 16-ways, conceptually increasing the associativity to 128. However, this design requires accessing 128 locations on each cache access to check if an address is in the cache or not, resulting in a high power overhead of 67\%~\cite{PhantomCache}. Moreover, this design is potentially vulnerable to future eviction set discovery algorithms as it selects a victim line from only a subset of the cache lines. In comparison, Mirage provides the security of a fully-associative cache where any eviction-set discovery is futile, with practical overheads.

{\bf HybCache}~\cite{Hybcache} is a recent design providing fully-associative mapping for a subset of the cache (1--3 ways), to make a subset of the processes that map their data to this cache region immune to eviction-set discovery. However, the authors state that ``applying such a design to an LLC or a large cache in general is expensive''~\cite{Hybcache}. For example, implementing a fully-associative mapping in 1 way of the LLC would require parallel access to >2000 locations per cache-lookup that would considerably increase the cache power and access latency). In contrast, Mirage provides security of a fully-associative design for the LLC with practical overheads, while accessing only 24--28 locations per lookup.

\subsection{Cache Associativity for Performance}
{\bf V-Way Cache}~\cite{Vway}, which is the inspiration for our design, also uses pointer-based indirection and extra tags to reduce set-conflicts -- but it does not eliminate them. V-Way Cache uses a set-associative tag-store, which means it is still vulnerable to set-conflict based attacks, identical to a traditional set-associative cache. Mirage builds on this design and incorporates skewed associativity and load-balancing skew-selection to ensure set-conflicts do not occur in system-lifetime. 

{\bf Z-Cache}~\cite{zcache} increases associativity by generating a larger pool of replacement candidates using a tag-store walk and performing a sequence of line-relocations to evict the best victim. However, this design still selects replacement candidates from a small number of resident lines (up to 64), limited by the number of relocations it can perform at a time. As a result, a few lines can still form an eviction set, which could potentially be learned by attacks. Whereas, Mirage selects victims globally from all lines in the cache, eliminating eviction-sets.

\ignore{
{\bf Z-Cache}~\cite{zcache} expands the ability of the cache to choose the replacement victim beyond a single set by relocating an evicted line to an alternate location if that location has a line with a lower value of the LRU timestamp. These relocations can be repeated multiple times to find a line with even lower value of the LRU timestamp.  However, such a design still selects the replacement candidates from only a small number of resident lines (up to 64), limited by the number of relocations that can be performed without impacting the cache miss latency. The limited set of replacement candidates means that a few lines can form an eviction set, which could potentially be learned by future algorithms. Whereas, Mirage selects the victim globally from all the data lines in the cache. 
}

{\bf Indirect Index Cache}~\cite{iic} is a fully-associative design that uses indirection to decouple the tag-store from data-blocks and has a tag-store designed as a hash-table with chaining to avoid  tag-conflicts. However, such a design introduces variable latency for cache-hits and hence is not secure. While Mirage also uses indirection, it leverages extra tags and power of 2 choices based load-balancing, to provide security by eliminating tag-conflicts and retaining constant hit latency.

\textbf{Cuckoo Directory}~\cite{cuckooDirectory} enables high associativity for cache-directories \ignore{with only a small number of ways (3--4), }by over-provisioning entries similar to our work and using cuckoo-hashing to reduce set-conflicts. \ignore{Similar ideas have also been adopted by }\textbf{SecDir}~\cite{isca19:secdir} also applies cuckoo-hashing to protect directories from conflict-based attacks~\cite{DirectoryAttacks}. However, cuckoo-hashing alone is insufficient for conflict-elimination. Such designs impose a limit on the maximum number of cuckoo relocations they attempt (e.g. 32), beyond which they still incur an SAE. In comparison, load-balancing skew selection, the primary mechanism for conflict-elimination in Mirage, is more robust at eliminating conflicts as it can ensure no SAE is likely to occur in system-lifetime with 75\% extra tags.





\subsection{\mbox{Isolation-based Defenses for Set-Conflicts}}

Isolation-based defenses attempt to preserve the victim lines in the cache and prevent conflicts with the attacker lines. Prior approaches have partitioned the cache by sets~\cite{MI6,PageColoring} or ways~\cite{RPCache,Catalyst,DAWG,CacheBar} to isolate security-critical processes from potential adversaries. However, such approaches result in sub-optimal usage of cache space and are unlikely to scale as the number of cores on a system grows (for example, 16-way cache for a 64-core system). Other mechanisms explicitly lock security-critical lines in the cache~\cite{RPCache,StealthMem}, or leverage hardware transactional memory~\cite{Cloak} or replacement policy~\cite{sharp} to preserve security-critical lines in the cache. However, such approaches require the classification of security-critical processes to be performed by the user or by the Operating-System.  In contrast to all these approaches, Mirage provides robust and low-overhead security through randomization and global evictions, without relying on partitioning or OS-intervention.

\vspace{-0.05in}
\section{Conclusion}
\vspace{-0.05in}
Shared LLCs are vulnerable to conflict-based attacks. Existing randomized LLC defenses continue to be broken with advances in eviction-set discovery algorithms. We propose Mirage as a principled defense against such attacks. Providing the illusion of a fully-associative cache with random-replacement, Mirage guarantees the eviction of a random line on every cache-fill  that leaks no address information, for  $10^4 - 10^{17}$ years. Mirage achieves this strong security 
with 2\% slowdown and modest area overhead of 17-20\%, compared to a non-secure set-associative LLC.
Thus, Mirage provides a considerable safeguard against current eviction-set discovery algorithms and potentially against even future advances.

\vspace{-0.1in}
\section*{Acknowledgments}
We thank Ananda Samajdar for help in setting up the RTL synthesis tool-chain. 
We also thank the anonymous reviewers and members of Memory Systems Lab, Georgia Tech for their feedback. This work was supported in part by SRC/DARPA Center for Research on Intelligent Storage and Processing-in-memory (CRISP) and a gift from Intel. Gururaj Saileshwar is partly supported by an IISP Cybersecurity PhD Fellowship.

\bibliographystyle{plain}
\bibliography{sigproc}

\begin{figure*}[thb] 
  	\centering
  \vspace{-0.425 in}
\includegraphics[width=6in]{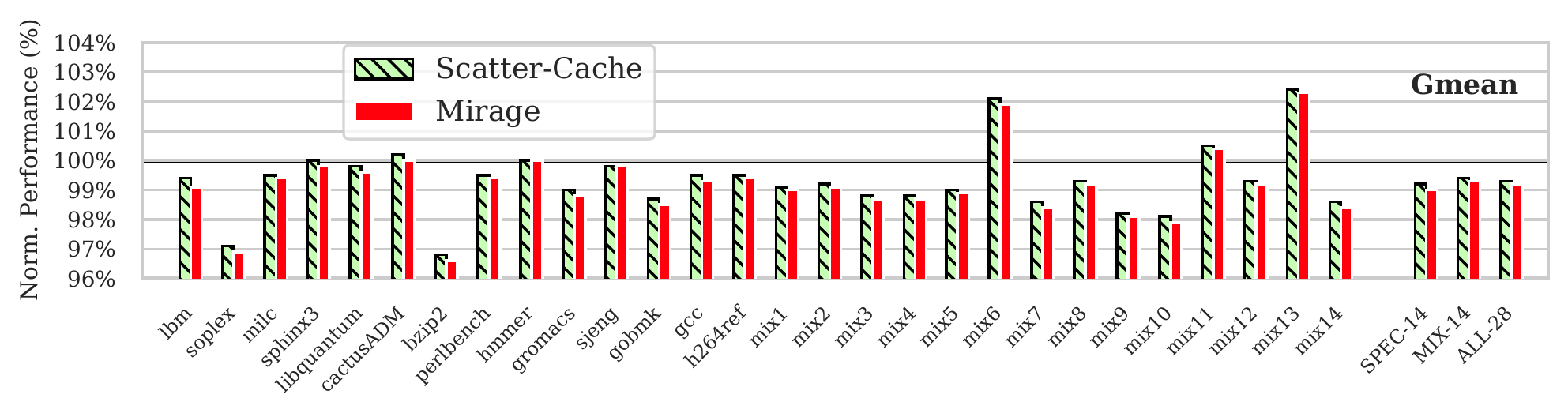}
 \vspace{-0.225in}
\caption{
Gem5-based performance evaluation. Performance of Mirage and Scatter-Cache normalized to Non-Secure Baseline (using the weighted speedup metric). On average, Mirage incurs a  slowdown of 0.8\%, and  Scatter-Cache of 0.7\%.
}
 \vspace{-0.15in}
	\label{fig:perf_gem5} 
	
\end{figure*}

\appendix
\begin{appendices}

\vspace{-0.05in}
\section{Validation with RISC-V RTL}
\label{sec:riscv_validation}
\vspace{-0.05in}

To validate our results with a hardware design, we implemented randomized caches in RISC-V hardware.
We use Firesim~\cite{Firesim}, the state-of-the-art platform for FPGA-based cycle-exact simulation of RISC-V cores on AWS FPGAs. Unfortunately, all RISC-V processors currently only support a two-level cache hierarchy by default. While FireSim emulates a last-level cache (L3 cache), it  only models the tag-store and not the data-store; the timing model on the FPGA is stalled until the data is functionally accessed from the host DRAM~\cite{FASED}. Without the data-store for the L3 Cache, it is infeasible to directly implement Mirage. However, as Mirage has a similar LLC-miss count as Scatter-Cache and a 1 cycle higher access latency (due to FPTR lookup), we can estimate its performance by implementing a randomized cache design with two-skews (similar to Scatter-Cache) and increasing the cache access latency by one cycle to account for the FPTR lookup. For implementing cache  randomization, we used a hardware implementation of 3-cycle PRINCE cipher.

We perform the study using a 4MB/16-way L3 cache (the default size of L3 in the FireSim 4-core Rocket-Core design).  Table~\ref{table:RISCV} compares execution time (in billion cycles) for a baseline set-associative LLC (Base) and the randomized cache design as the lookup latency of the cache is increased by 3 cycles to 6 cycles.  Note that for this evaluation, we run the  SPEC2017-Int workloads to completion. On average, the randomized cache design with even six cycle additional lookup latency causes only a 1\% slowdown on average. Thus, the slowdown from the RISC-V FPGA-based evaluation is quite similar to the slowdown from our simulator (2\%). 

\begin{table}[htb]
\vspace{-0.1in}
  \begin{center}
   \renewcommand{\arraystretch}{1.15}
     \setlength{\tabcolsep}{4pt}
    \caption{Execution time (in billion cycles) of RISC-V for Non-Secure LLC (Base) and randomized cache where the cache lookup latency is increased by 3 to 6 cycles.}
  \vspace{0.1in}
    \begin{small}
    \resizebox{3.3in}{!}{
      \begin{tabular}{|l||c|c|c|c|c|} \hline
      
    \multirow{2}{*}{\textbf{Workload}}   & \multirow{2}{*}{\textbf{Base}} & \multicolumn{4}{c|}{\textbf{Randomized cache with increased lookup latency}}  \\ \cline{3-6}
                                & & \textbf{+3 cycles} & \textbf{+4 cycles} & \textbf{+5 cycles} & \textbf{+6 cycles}\\ \hline \hline
            perlbench  &   191 &    202  &    194  &    206    &  203      \\ \hline
            mcf        &   191 &    199  &    194  &    200    &  201      \\ \hline
            omnetpp    &    42 &    42   &   41    &     42    &  42       \\ \hline
            x264       &   699 &    707  &    702  &    696    &  707      \\ \hline
            deepsjeng  &    85 &    84   &   84    &     84    &  84       \\ \hline
            leela      &    44 &    44   &   45    &     45    &  45       \\ \hline
            exchange2  &   109 &    110  &    108  &    108    &  109      \\ \hline
            xz         &   119 &    114  &    114  &    115    &  115      \\ \hline \hline
            MEAN       &  100\% &  100.6\% &   99.5\% &   100.9\%  &  101.0\%   \\ \hline
      \end{tabular}
      }
    \end{small}   
\vspace{-0.3in}
\label{table:RISCV}
\end{center}
\end{table}

\vspace{-0.05in}
\section{Validation with Gem5 Simulator}
\label{sec:gem5_validation}
\vspace{-0.05in}

We also validated our simulator results using Gem5~\cite{Gem5}, a cycle-accurate micro-architecture simulator. As the default implementation of Gem5 does not support a 3-level cache hierarchy, which is typical in modern processors,
 we did not pick Gem5 for evaluations in our paper. However, for reproducibility, we re-implemented Mirage and Scatter-Cache in Gem5\footnote{The artifact-evaluated Gem5 implementation of Mirage is available open-source at \url{http://github.com/gururaj-s/MIRAGE}.} for the L2 cache (in the Gem5 2-level cache hierarchy) and validated that all the misses in Mirage result in Global Evictions (no SAE). Figure~\ref{fig:perf_gem5} shows the performance of Scatter-Cache (SC) and Mirage normalized to a non-secure set-associative LLC baseline for a 4-core system with an 8MB L2 cache as the LLC running SPEC-CPU2006 workloads (simulated for 1 billion instructions after forwarding the first 10 billion instructions). Averaged across 14 memory-intensive SPEC workloads (4 copies of a benchmark on 4 cores) and 14 mixed workloads (random combinations of 4 benchmarks), Mirage incurs a slowdown of 0.8\% while SC incurs a slowdown of 0.7\%,  within our simulator results of 2\% and 1.7\% 
slowdown
respectively. 
\vspace{-0.15in}
\section{Efficacy of Load-Aware Selection}
\vspace{-0.075in}

We provide intuition with the buckets and balls model (buckets equivalent to cache-sets and balls equivalent to cache-installs) using bounds from Mitzenmacher's thesis~\cite{MitzenmacherRealThesis}. Consider $N$-balls thrown in $N$-buckets ($avg\text{-}bucket\text{-}load=1$).  
With one skew (each ball maps to one random bucket), the non-uniformity in mapping causes some buckets to have higher load (most-loaded bucket has $O(log(N))$ balls). With two skews, a ball can go to two places, but the random skew selection has no intelligence in placement, i.e. a ball can end up in a bucket with high-load (the most-loaded bucket still has $O(log(N))$ balls). With 2 skews and load-aware skew-selection, a ball can go to two places and the placement specifically avoids the high-load bucket, thus reducing imbalance; this has been shown to reduce the most-loaded bucket load to $O(log(log(N)))$ balls. The gain from $O(log(N))$ to $O(log(log(N)))$ is dramatic, but going beyond 2 skews has diminishing returns as $log(log(N))$ already has little variation as $N$ changes; so we restrict our study of Mirage to 2 skews.


\end{appendices}


\end{document}